\documentclass[amsmath, amssymb, english, aps, prl, showpacs, twocolumn, reprint, floatfix,superscriptaddress]{revtex4-1}
\usepackage[pdftex,plainpages=false,colorlinks=true,linkcolor=blue, citecolor=blue, urlcolor=blue]{hyperref}
\usepackage{color}
\usepackage{amssymb}
\usepackage{epsfig}
\usepackage{graphicx}
\usepackage{amsmath}
\usepackage{array,color}
\usepackage{natbib}

\usepackage{footmisc}

\newcommand{\bk}{\mathbf{k}}
\newcommand{\bq}{\mathbf{q}}
\newcommand{\br}{\mathbf{r}}
\newcommand{\re}{\mathrm{Re}}
\newcommand{\im}{\mathrm{Im}}

\begin{document}

%\templatetype{pnasresearcharticle} % Choose template 
% {pnasresearcharticle} = Template for a two-column research article
% {pnasmathematics} %= Template for a one-column mathematics article
% {pnasinvited} %= Template for a PNAS invited submission

\title{Non-Fermi liquid phase and linear-in-temperature scattering rate in overdoped   two dimensional Hubbard model}

% Use letters for affiliations, numbers to show equal authorship (if applicable) and to indicate the corresponding author
%\author[a,*]{W\'ei W\'u}
%\author[a]{ Xiang Wang} 
%\author[b]{André-Marie Tremblay}

%\affil[a]{School of Physics, Sun Yat-sen University, Guangzhou, 510275, Guangdong, China}
%\affil[b]{D\'epartement de physique, Institut quantique and RQMP, Universit\'e de Sherbrooke, Sherbrooke, Qu\'ebec, J1K 2R1, Canada}

\author{Wei Wu}
\email[]{wuwei69@mail.sysu.edu.cn}

\affiliation{School of Physics, Sun Yat-sen University, Guangzhou, Guangdong Province 510275, China
}

\author{Xiang Wang}
\affiliation{School of Physics, Sun Yat-sen University, Guangzhou, Guangdong Province 510275, China
}

\author{André-Marie Tremblay}
\affiliation{D\'epartement de physique, Institut quantique and RQMP, Universit\'e de Sherbrooke, Sherbrooke, Qu\'ebec, J1K 2R1, Canada}

\begin{abstract}
Understanding electronic properties that violate the Landau Fermi liquid paradigm in cuprate superconductors remains a major challenge in condensed matter  physics. The strange metal state in overdoped cuprates that exhibits linear-in-temperature scattering rate and dc resistivity is a particularly puzzling example. Here, we compute the electronic scattering rate in the two-dimensional Hubbard model using cluster generalization of dynamical mean-field theory. We present a global phase diagram documenting an apparent non-Fermi liquid phase, in between the pseudogap and Fermi liquid phase in the doped Mott insulator regime. We discover that in this non-Fermi liquid phase, the electronic scattering rate  $\gamma_k(T)$  can display linear temperature dependence as temperature $T$ goes to zero.  In the temperature range that we can access, the $T-$ dependent  scattering rate is isotropic on the Fermi surface, in agreement with recent experiments.  Using fluctuation diagnostic techniques, we identify antiferromagnetic fluctuations as the physical origin of the $T-$ linear electronic scattering rate.
\end{abstract}

%\dates{This manuscript was compiled on \today}
%\doi{\url{www.pnas.org/cgi/doi/10.1073/pnas.XXXXXXXXXX}}

\maketitle
%\thispagestyle{firststyle}
%\ifthenelse{\boolean{shortarticle}}{\ifthenelse{\boolean{singlecolumn}}{\abscontentformatted}{\abscontent}}{}

% If your first paragraph (i.e. with the \dropcap) contains a list environment (quote, quotation, theorem, definition, enumerate, itemize...), the line after the list may have some extra indentation. If this is the case, add \parshape=0 to the end of the list environment.
%\dropcap{T}
\section*{Introduction}
The non-Fermi liquid states emerging from strongly correlated electron systems have been one of the central research topics in condensed matter physics~\cite{Stewart2001}.
One of the most profound problems in this field is the strange metal state in cuprates, characterized by a linear temperature dependence of dc resistivity,  and a  scattering rate $1/\tau$  reaching a putative universal ``Planckian limit",  $ \hbar / \tau = k_B T $
~\cite{Legros2019,zaanen2019,Shen2020,Varma2020,hartnoll2021planckian,Ayres_Berben_Hussey_2021}.
Since the discovery of strange metallicity in cuprates~\cite{cooper2009,Daou2009,Hussey2011} and 
other materials~\cite{lohenysen1994,Grigera329, DoironL2009, Shen2020}, enormous effort has been aimed at tracing its physical origin, including phenomenological theories~\cite{mfl,Rice2017}, considerations on  quantum critical fluctuations in vicinity of a quantum critical point (QCP)~\cite{ millis1993,abanov2003, gegenwart2008,  lohneysen2007, Xu2020,dumitrescu2021, Cha18341},  and also studies of microscopic models~\cite{sachdev93,patel2019} in the absence of a nearby QCP, such as the Sachdev-Ye-Kitaev (SYK) type models with random interactions~\cite{sachdev93,parcollet1999,patel2019}. Up to date,  however, 
the rigorous relevance  of these models  to overdoped cuprates is still far from clear, since little is known about the underlying mechanism of the strange metal state.

The two-dimensional Hubbard model, which is  prevalent in modeling correlated materials,  can capture various signature features of hole-doped cuprates, such as d-wave superconductivity~\cite{scalapino2007,maier_d:2005,gull_superconductivity_2013,fratino2016organizing}, pseudogap~\cite{macridin2006,senechal2004,sordi:2012,wu2018,Reymbaut:2019},  stripe order~\cite{Zheng1155,Dash_Senechal_2020}. 
%However, a systematic study of its non-Fermi liquid properties in the overdoped regime is still lacking.
Recently, in studies  at very high temperatures ($T\sim $ bandwidth $W$), the so-called ''bad metal'' regime of the Hubbard model  has  been reported~\cite{perepelitskey2016, huang2019strange,cha2020,Brown_Mitra_Bakr_2018}. In those studies, the high temperature $T-$ linear resistivity stems  largely from  a change in effective carrier number with temperature~\cite{Gunnarsson2003, cha2020}.  This is in stark contrast to  cuprate materials, where the  $T-$ linear dc resistivity occurs at low temperature, the so-called ``strange metal'' regime. In this regime, it is argued that linear-in-temperature resistivity originates from a scattering rate $1/\tau$ that scales linearly with temperature~\cite{grissonnanche2020} and reaches a putative fundamental limit set by ''Planckian dissipation''~\cite{zaanen2019}. Whether the Hubbard model can provide a proper description of  the cuprate strange metal at low temperatures  is therefore still 
a crucial open question.
  
To address these problems,  in this work we  solve the two dimensional Hubbard at low temperatures on a square lattice, in the doped Mott-insulator regime using the dynamical cluster approximation (DCA)~\cite{maier2005}. We demonstrate that the  $T-$  linear  electronic scattering rate at low temperatures, found in the strange metal state of hole-doped cuprates~\cite{Legros2019,Chen1099}, can emerge from  the overdoped Hubbard model. The inelastic part of the $T-$  linear  electronic scattering rate is the same at the node and at the antinode. Our results suggests that although the scattering rate is close to the Planckian one, that rate does not seem to be a limit for reasons that we explain.   More importantly, we explicitly identify that the short-ranged antiferromagnetic correlations, despite being greatly suppressed in the overdoped regime,  are at the origin of the   $T-$  linear scattering rate characterizing strange metallicity.

 We consider the Hubbard model Hamiltonian,
 \begin{equation}
 \mathcal{H} = \sum_{ij,\sigma} -t_{ij} c^{\dagger}_{i,\sigma} c_{j,\sigma} + U \sum_{i} n_{i\uparrow} n_{i\downarrow} - \mu \sum_{i,\sigma} n_{i\sigma},
 \end{equation}
where  $\mu$ is the chemical potential, 
the $t_{ij}$'s are non-zero for nearest-neighbor hoppings $t$,  and
next-nearest-neighbor hoppings $t'$, which varies in different cuprate compounds~\cite{pavarini2001}. $U$ is the onsite Coulomb repulsion, which is taken as $U/t=7$ through out this work. We work in units where $t=1$, the lattice spacing, Boltzmann's constant $k_B$ and Planck's constant $\hbar$ are also set to equal to unity.
The DCA method is a  cluster extension of the dynamical mean-field theory (DMFT)~\cite{georges1996review}  that treats quantum and short-ranged spatial correlations exactly,
 while longer range correlations beyond the cluster are incorporated  in a dynamical mean-field way (see Materials and Methods).

\section*{Results}

\begin{figure}
\centering
\includegraphics[width=.8\linewidth]{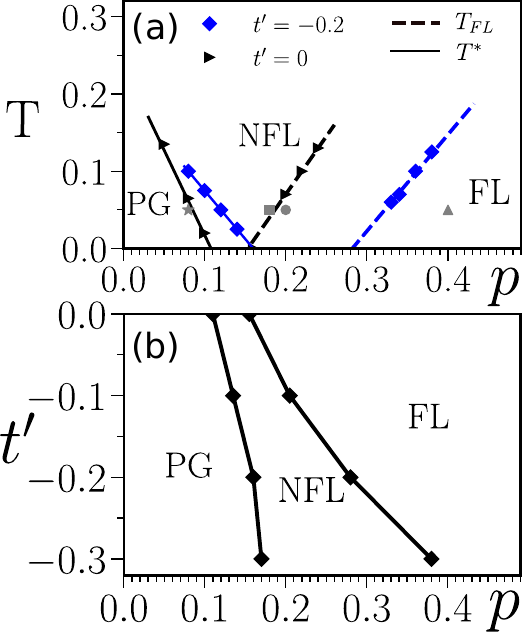}
  \caption{
    \textbf{Pseudogap (PG), Non Fermi liquid (NFL)  and Fermi liquid  (FL) phases of the doped Hubbard model in normal state.}
    \textbf{(a):} Pseudogap temperature $T^{*}$ and Fermi liquid temperature $T_{FL}$  as a  function of hole doping value $p$ for two typical $t^{\prime}$ values, $t^{\prime} = 0$ (Triangles)  and $t^{\prime} = -0.2$ (Diamonds). 
    The finite temperature data points are extrapolated
    to zero temperature (lines),  yielding two critical dopings $p^*$ and
    $p_{FL}$. For example, for $t^{\prime} = -0.2$, $p^* \simeq 0.16$ and $p_{FL} \simeq 0.28$. For the definition of $T^{*}$ and $T_{FL}$ please refer to main text  and Supplementary Sec.~\ref{Sup:T*TFL}.
    The gray symbols mark the data points that are further analyzed in Fig.~\ref{fig:fluc} at $t^{\prime} = -0.2t$. 
    \textbf{(b):} Zero temperature phase diagram in the  $p-t^{\prime}$ plane. The above extrapolated $p^*$ and  $p_{FL} $ define the PG/NFL and NFL/FL phase boundaries respectively. 
   } \label{fig:phase}
\end{figure}

\subsection*{Phase diagram} We first  display two characteristic energy scales of the doped normal state Hubbard model: the pseudogap temperature $T^{*}$, and the Fermi liquid temperature $T_{FL}$,  as a function of doping levels $p$ in Fig.~\ref{fig:phase}a. Here $T^{*}$ is defined as the temperature $T$ where the antinodal zero-frequency
spectral function starts to decrease with $T$, and  $T_{FL}$ is identified as the temperature where the paramagnetic susceptibility  (Knight shift) becomes $T$-independent (see Supplementary Fig.\ref{fig:tt}).  
%The definitions of $T^{*}$ and $T_{FL}$ can be find in the caption of Fig.~\ref{fig:phase} and the supplementary.
 Extrapolating  $T^{*}$ and $T_{FL}$ to zero, one finds two critical dopings: $p^{*}$ where pseudogap disappears for $p> p^{*}$, and $p_{FL}$ where  
Fermi liquid emerges for  $p> p_{FL}$.
 Repeating this calculation for several $t^{\prime}$ values, we obtained a zero temperature
   phase diagram  in the $p-t^{\prime}$ plane, as shown in Fig.1b, which consists of three different phases: 
 (1) PG phase in the underdoped regime  where  $p<p^{*}$  ( and $T^{*} > 0$). (2) Canonical FL phase on the heavily overdoped side  for  $p>p_{FL}$ (where $T_{FL} >0$).   (3) Finally, in between  PG and FL phases, there exists
a NFL phase  where the extrapolated $T^{*}$ and  $T_{FL}$ both vanish in the $p^{*}<p<p_{FL}$ interval.  Namely in the NFL phase,  there is no pseudogap at the Fermi level but the physical properties disagree with expectations for a Fermi liquid. 
It is remarkable that for all the $t^{\prime}$ values we have studied, the NFL resides in 
a finite range of dopings. In fact, as the value of $|t^{\prime}/t|$ increases, the NFL regime becomes broader in doping, as one can see from Fig.1a.
This result suggests that upon hole doping,  the pseudogap state does not directly transit to the  Fermi liquid phase via a  single quantum critical point at zero temperature. 

Comparing with experiments, we note that in $\mathrm{La_{2-x}Sr_{x}CuO_{4} (LSCO)}$ compound ($t^{\prime}/t \sim -0.2$), it is found that the PG ends at $p^{*} \simeq 0.18$,  and Fermi liquid shows up at $p_{FL} \simeq 0.3 $ [where $p_{FL} $ is  defined as where the temperature-dependent resistivity becomes $\rho(T) \propto T^{2} $ ~\cite{ cooper2009, greven2013}]. This is in good agreement with our result that the NFL exists  in the doping range $p\in (p^* = 0.16, p_{FL}= 0.28)$ at $t^{\prime}/t = -0.2$.
Recall that here the spontaneous symmetry breaking phases,  such as  the d-wave superconductivity (SC),  are suppressed to simulate transport experiments in high magnetic field.

\begin{figure*}
\centering
    \includegraphics[width=1.8\columnwidth]{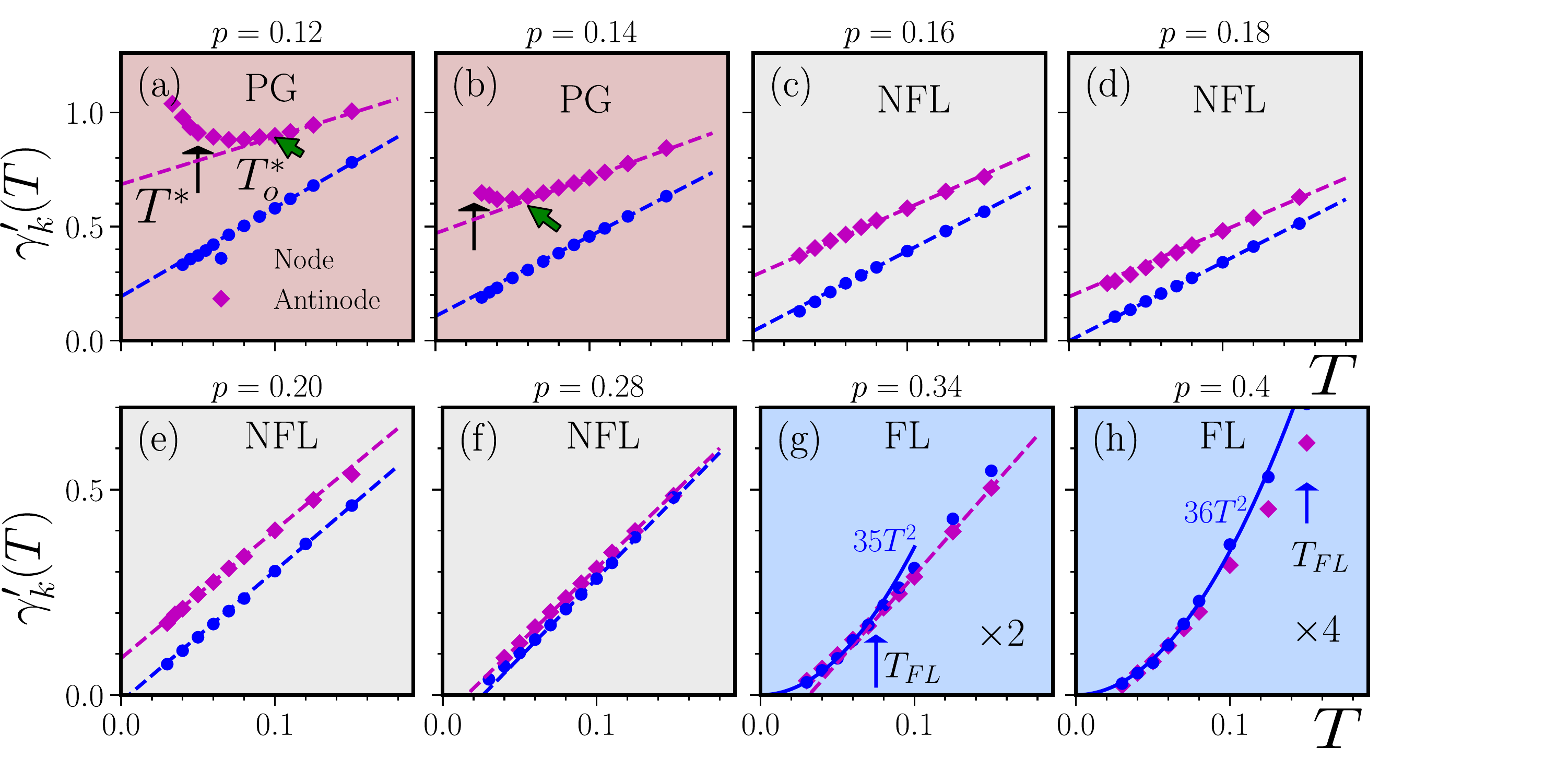}
  \caption{
    \textbf{Temperature dependence of the electron  scattering rate.}
 Here $\gamma^{\prime}_k\equiv -\im \Sigma ^{(2)} (k,\omega=0)$ is shown as a function of temperature $T$ for different dopings. Dashed lines show linear fittings $\gamma^{\prime}_k= aT+b$,  while solid lines show quadratic fittings $\gamma^{\prime}_k= aT^2$. For example, at $p=0.18$, for antinode  $\gamma^{\prime}_k\approx 3.13T+0.17$ and for node $\gamma^{\prime}_k\approx 3.45T$ , while $\gamma^{\prime}_k\approx 9T^2$ for nodal $\gamma^{\prime}_k$ at $p=0.4$. Note that  in the last two subplots, $\gamma^{\prime}_k$ data is enlarged for clarity. The pseudogap temperature $T^{*}$
 , and the temperature $T^{*}_{o}$ where  $\gamma^{\prime}_k$ starts to deviate from linearity are marked by  arrows in subplots (a) and (b). In subplots (g) and (h) the Fermi liquid temperature $(T_{FL})$ is also indicated by arrows. For the definition of $T^{*}$ and $T_{FL}$, please refer to main text and Supplementary Sec.~\ref{Sup:T*TFL}. 
 }
\label{fig:scatter}    
\end{figure*}

\subsection*{$T-$ linear scattering rate} The electronic scattering rate  $\gamma_k \equiv -\mathrm{Im}\Sigma(k, \omega=0)$ in the NFL phase is the primary focus of this work. 
%At this point, we do not make any quasiparticle assumption. 
We find that in the NFL, the Matsubara data for the self-energy  $\Sigma(\bk, i\omega_n)$ is consistent with the hypothesis that the imaginary part of the self-energy in real frequency space $\Sigma^{''}(k,\omega) \equiv \mathrm{Im} \Sigma(k, \omega)$ follows an $\omega/T$ scaling~\cite{parcollet1999,mfl,schroder2000onset,schafer} at low-energies (see Supplementary Fig.~\ref{fig:ken}-\ref{fig:scale}).
 Hence we assume that $\Sigma^{''}(k,\omega)$ can be written as $\Sigma^{''}(k,\omega)=-T^{\alpha}\phi(\omega/T)-b$~\cite{schafer,Chen1099} at low-energies,  where $\phi(\omega/T)$ is an analytic function of $\omega/T$, while $\alpha$ and $b$ are constants. With this assumption, the imaginary part of the self-energy at zero-frequency that follows from a second order polynomial extrapolation in Matsubara frequencies $\gamma^{\prime}_k\equiv -\mathrm{Im}\Sigma^{(2)}(k, \omega=0) =-\im [1.875\Sigma(k,   i\omega_0)-1.25 \Sigma(k, i\omega_1)+ 0.375 \Sigma(k,  i\omega_2)]  = aT^{\alpha}+b $ will have exactly the same $T-$ dependence of the true scattering rate $\gamma_k $,  since the scaling hypothesis implies that $\gamma_k = \phi(0)T^{\alpha}+b$. Therefore, one can find the \textit{ exact} exponent $\alpha$ describing the $T-$ dependence of $\gamma_k $ from analyzing the $ \gamma^{\prime}_k$ data, despite the fact that the fit leaves the constant coefficient  $\phi(0)$  unknown
  [ if  $\Sigma^{''}(k,\omega)$ is $\omega-$ independent over the frequency range $| \omega |  \lesssim 4T$, $ \phi(0)  \approx a$, see Supplementary Sec.\ref{sup:omega} for details].

Throughout the following, we use the typical value $t^{\prime} = -0.2$  as an example to study the $T-$ linear scattering rate. Fig.~\ref{fig:scatter} displays  $\gamma^{\prime}_k$ as  a function of temperature $T$ for different $p$ values, where one can see that at high temperatures, the scattering rate $\gamma^{\prime}_k$  is linear in temperature in a  remarkablely large doping range, from underdoped  ($p=0.12$, Fig.~\ref{fig:scatter}a ) to heavily overdoped side ($p=0.34$, Fig.~\ref{fig:scatter}g )~\cite{greven2013}.

When $T$ is decreased, focusing on the aninodal $\gamma^{\prime}_k$ at $\bk=(0,\pi)$ as shown in Fig.~\ref{fig:scatter}a-b,
 at small dopings ($p=0.12, 0.14$, in the PG),  $\gamma^{\prime}_k$  deviates from its high temperature  $T-$ linearity, developing a prominent upturn when  the pseudogap temperature $T^{*}$ is reached. This resembles the upturn seen  in the dc resistivity curves in transport experiments~\cite{taillefer2018}, and  in other calculations~\cite{GullFerrero:2010,SordiResistivity:2013}, which characterizes the opening of pseudogap. 
 As the doping level $p$ increases,  the upturn of  $\gamma^{\prime}_k$  at the antinode  shifts to lower temperatures in the the PG phase, reflecting the decreasing $T^*$.
%As one can  see from Fig.2a-c that as doping level $p$ is increased, the temperature where   $\gamma^{\prime}_k$ deviates linearity (marked as $T^{*}_o$ in Fig. 2a-b) decreases along with the decreasing  pseudogap temperature $T^{*}$. Finally,  $\gamma^{\prime}_k$ at antinode recovers its perfect linear $T-$ dependence in the full temperature regime as shown in Fig.2c, when NFL phase is reached. 
Finally, when the NFL phase is reached, a possible upturn of $\gamma^{\prime}_k$ moves outside of the acessible temperature range. The linear $T-$ dependence of $\gamma^{\prime}_k$ at the antinode  extends to $T\rightarrow 0$ ,  as shown in Fig.~\ref{fig:scatter}d.
For the node,   $\gamma^{\prime}_k$ preserves the linear- in-$T$ behavior,  crossing  the PG -NFL transition. Thus for a typical doping close to $p^*$ in the NFL,  $p=0.18$ (Fig.~\ref{fig:scatter}d) for example, both the node and the antinode display a $T-$ linear scattering rate in the full temperature regime.
On the  heavily overdoped side, the  Landau Fermi liquid paradigm is restored at small $T$ when $p>p_{FL}$. As shown in Fig.~\ref{fig:scatter} g-h,  the scattering rates crossover from  high $T-$ linearity  to a clear $T-$ square behavior ~\cite{xu2013} as $T < T_{FL}$. 
% Meanwhile,  the node-antinode differentiation in scattering rate becomes negligible.

In essence,  at low dopings $\gamma^{\prime}_k$ has upturns that characterize the PG,  while at large dopings it follows the  $T^2$  law that characterizes the FL.  In the NFL, where  $T^{*}$ and $T_{FL}$  are both vanishingly small,  $\gamma^{\prime}_k(T)$ obeys $\gamma^{\prime}_k(T) =aT+b$  in a broad $T$ range. 
Nevertheless, we point out that in the NFL, when doping $p$ is close to $p^{*}$ or $p_{FL}$, the precursor effects of pseudogap or Fermi liquid  at  small $T$ can also break the  $T-$  linearity of $\gamma^{\prime}_k(T)$ , even if  $T^{*}$ or $T_{FL}$ appear to vanish (see Supplementary Fig.~\ref{fig:allanti}). As a result,  in the $T \rightarrow 0 $ limit , $\gamma^{\prime}_k(T)=aT+b $  is obeyed only in a part of the NFL regime.  For example, at $t^{\prime}/t = -0.2$, while  our definition suggests that the NFL exists in $ 0.16 \lesssim p \lesssim 0.28$ at vanishing $T$ (see discussions in Supplementary Sec.\ref{sup:tlinear}), the perfect linear-in-$T$ behavior of  $\gamma^{\prime}_k(T)$ [ or equivalently  the  linear-in-$T$ behavior of $\gamma_k(T)$ ] occurs in the doping range of $ 0.17 \lesssim p \lesssim 0.20$  as $T\rightarrow 0$,

Up to now, we have investigated  the electron scattering rate $\gamma_k=-\im\Sigma(k,\omega=0)$. In the FL regime, this differs from the quasiparticle scattering rate by a temperature-independent quasiparticle weight $z_k$.  In the NFL regime, it is worthwhile to investigate the phenomenological marginal Fermi liquid (MFL) interpretation of the scattering rate $1/\tau_k = z_k \gamma_k$ with $\gamma_k=-\im \Sigma(\bk,\omega)= \alpha \mathrm{max}(|\omega|, \pi T)+b$~\cite{mfl, Varma2020}. The procedure for finding $\tau_k$ from fitting the Matsubara Green's function is explained in Supplementary Sec.~\ref{sup:quasi}.  
We find $1/\tau_k  \sim CT $, with $C \in (1\sim2)$ (see  supplementary Fig.~\ref{fig:fit}-\ref{fig:mem}) for two doping levels, $p=0.18$ and $p=0.2$, in the $T-$ linear regime. We stress that here $C$ is found dependent on doping $p$ and momentum $\bk$. It decreases  as $p$ increases, contrary to what we found for the electron-scattering rate, which is nearly independent of doping in the NFL regime. 
%In addition, the marginal Fermi liquid scattering rate at $p=0.20$ differs in the antinodal and nodal directions, contrary to the electron scattering rate.  

\subsection*{Origin of the NFL and $T-$ linearity}

\begin{figure}
\centering
    \includegraphics[width=\columnwidth]{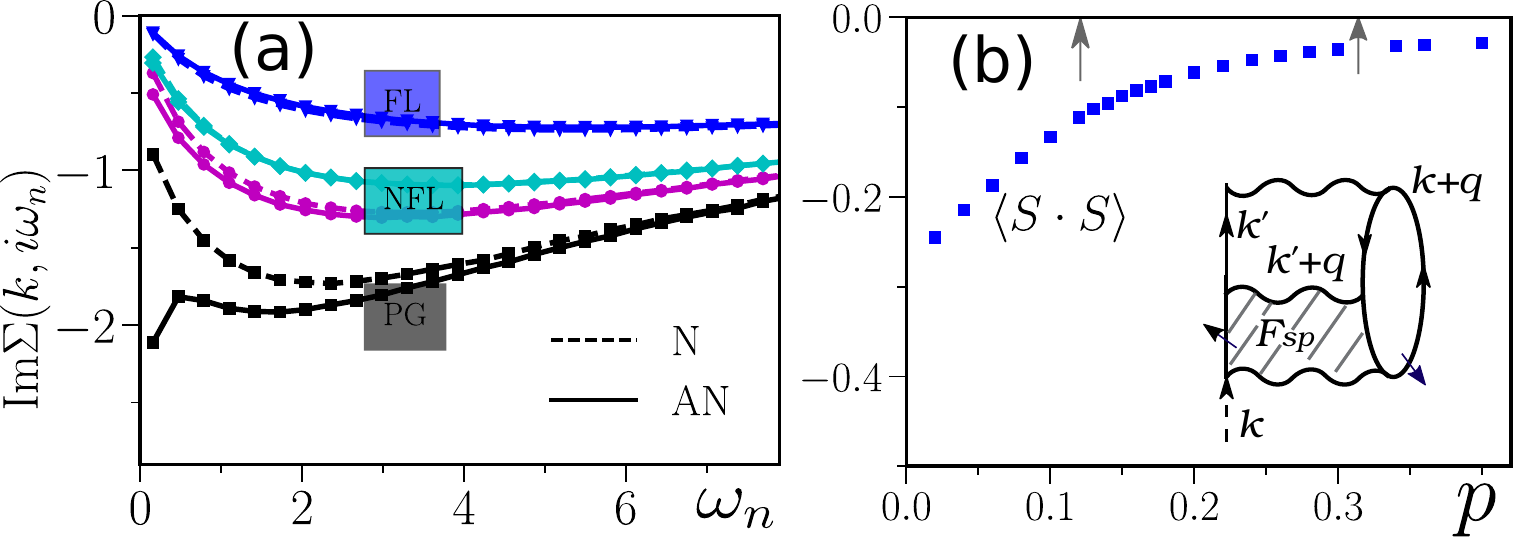}
  \caption{
    \textbf{Imaginary part of the self-energy in PG, NFL, FL phases,  and the spin-spin correlator $\langle S_{i}\cdot S_{i+1} \rangle$ on nearest neighboring sites $\langle i, i+1 \rangle$ as a function of doping $p$.}
    \textbf{(a):} $\im\Sigma(k,i\omega_n)$ as a function of $\omega_n$ . From bottom to top: $p=0.08$ (PG), $p=0.18$, $p=0.24$ (NFL),  and $p=0.4$ (FL). 
    \textbf{(b):} Spin correlator $\langle S_i \cdot S_{i+1} \rangle$ at two neighboring sites ($i, i+1$) as a function of doping $p$. Two arrows indicate PG/NFL and NFL/FL phase boundaries respectively. 
     \textbf{(Insert):} Feynman diagram that sketches the Dyson-Schwinger equation of motion decomposition (DSEOM) of the self-energy in the spin channel. 
     Here $U=7t, t^{\prime} = -0.2t, T=0.05t$.
   } \label{fig:sigma}
\end{figure}

%Recent theoretical efforts in  explaining the $T-$ linearity in  cuprates generally assume the strange metal  physics is  local in space, which intuitively seems  plausible,  since strange metal roots in the overdoped side of cuprates, where large doping can significantly reduce spatial correlations. However, this hypothesis has not yet been strictly rationalized by theoretical consideration nor by experimental verification. 
To reveal the physical origin of the $T-$ linear scattering rate in overdoped Hubbard model,
we use the  Dyson-Schwinger equation of motion (DSEOM) to decompose the self-energy at the two-particle level~\cite{gunnarsson2015, wu2017}.  Simply explained, the essential idea of this approach is to find how collective modes in different channels [\textit{spin (sp)},  \textit{charge (ch)} or \textit{particle-particle (pp)}] contribute to the self-energy.  As depicted by the  Feynman diagram for the spin channel in the insert of Fig.~\ref{fig:sigma}b,   the self-energy (with Hatree term  ${Un}/{2}$ subtracted) can be written as~\cite{gunnarsson2015}, 
\begin{equation}
\Sigma(k)- \frac{Un}{2}=-\frac{U}{g(k)\beta^2N}\sum_{k^{\prime},Q}F_{sp}(k,k^{\prime},Q)g(k^{\prime})g(k)g(k^{\prime}+Q)g(k+Q)
\end{equation}
where wave vectors $k$ stand for $k=(\bk, i\omega_n)$ and $g(k)$ is the full single particle Green's function. Here  $F_{sp}$ is the full two-particle scattering amplitude in the transverse spin channel. Hence the right-hand side of the above equation can be  rewritten in terms of the spin operators   $S^{+}_{k}(-Q)= C_{k\uparrow}^{\dagger}C_{k+Q,\downarrow}$, and $S^{-}_{k^{\prime}}(Q)= C_{k^{\prime}+Q\downarrow}^{\dagger}C_{k^{\prime},\uparrow}$,
\begin{eqnarray}
\Sigma(k)=\frac{-U}{g(k)\beta^2 N}\sum_{k^{\prime},Q} \langle S^{+}_{k}(-Q) S^{-}_{k^{\prime}}(Q) \rangle
\end{eqnarray}
 and we can  introduce a new quantity $\Sigma^{Q}_{sp}(k)=[-U/g(k)]\sum_{k^{\prime}} \langle S^{+}_{k}(-Q) S^{-}_{k^{\prime}}(Q) \rangle $
such that $\Sigma(k)=\sum_Q\Sigma^{Q}_{sp}(k) $, which has a clear physical meaning: the ratio  $|\im\Sigma^{Q}_{sp}(k)/\im\Sigma(k)|$  tracks the relative importance of the spin excitation with the momentum/frequency  transfer $Q$ to  the electronic scattering. The above analysis can also be straightforwardly applied to \textit{charge} and \textit{particle-particle} representations to estimate the impacts of the corresponding two-particle excitations on the self-energy (see Supplementary Sec.\ref{sup:fluc}).

\begin{figure}
\centering
    \includegraphics[width=0.99\columnwidth]{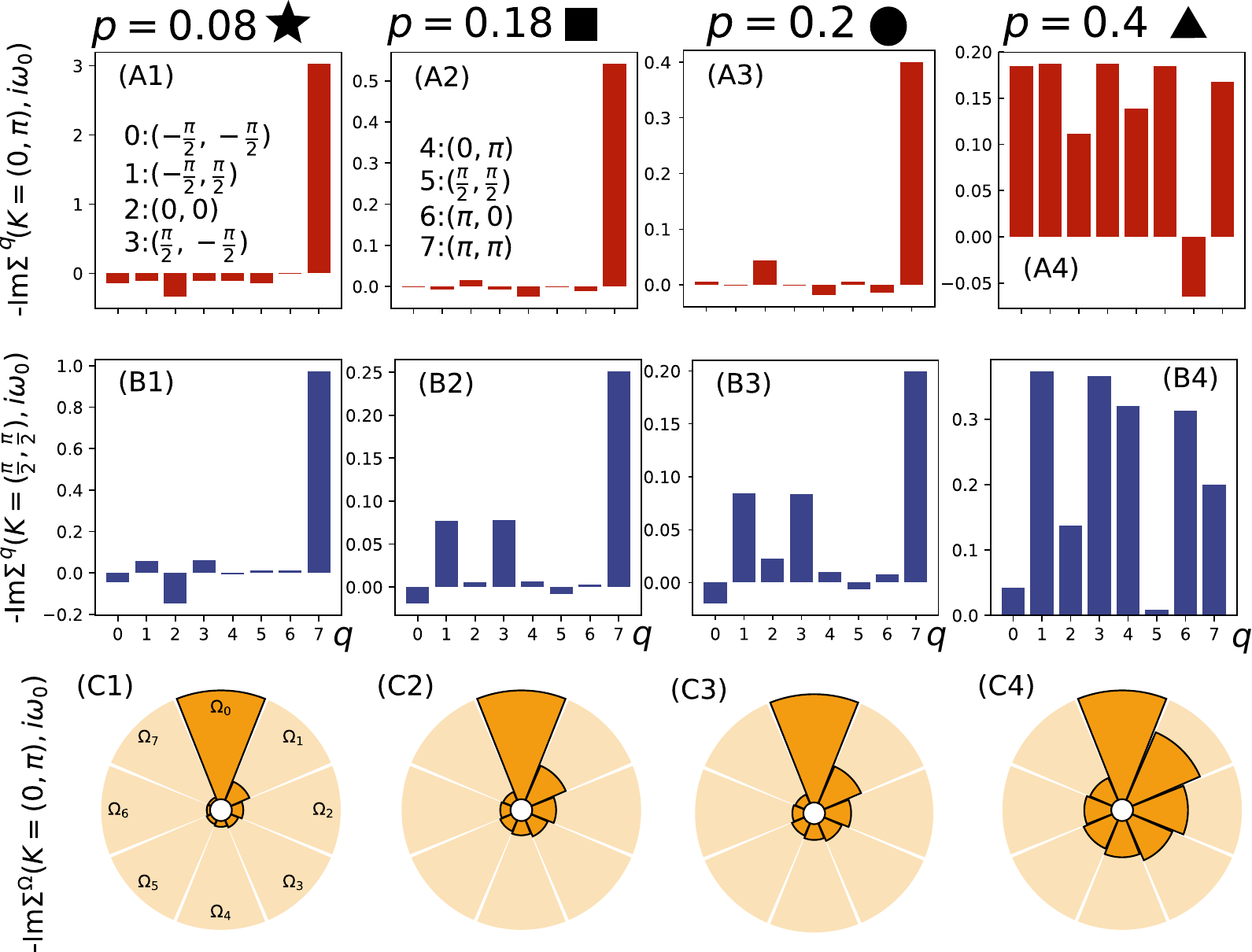}
  \caption{
    \textbf{Dyson-Schwinger equation of motion decomposition of the self-energy in \textit{spin} channel, $\im\Sigma^{\bq/\Omega_n}_{sp}(\bk, i\omega_0)$ at different dopings. }
    \textbf{(A1-A4):} $\im\Sigma^{\bq}_{sp}(\bk, i\omega_0)$ as a function of transfer momentum $\bq$ for the antinode [$\bk=(0,\pi)$].
    \textbf{(B1-B4):} $\im\Sigma^{\bq}_{sp}(\bk, i\omega_0)$ as a function of transfer momentum $\bq$ for the node [$\bk=(\pi /2,\pi /2)$].
 \textbf{(C1-C4):}  $\im\Sigma^{\Omega_n}_{sp}(k, i\omega_0)$ as a function of transfer frequencies $\Omega_n=2n\pi T$ for the antinode [$\bk=(0, \pi)$].
 Values  of  indiced transfer momenta $\bq$ are labeled in A1 and A2.  $\im\Sigma^{\Omega_n}_{sp}(k, i\omega_0)$ for the node, not shown here, is similar to that of the antinode [C1-C4].
 }
\label{fig:fluc}    
\end{figure}

 In Fig.~\ref{fig:sigma}~a, $\im\Sigma(\bk,i\omega_n)$ is shown as a function of $\omega_n$  in  different states. Focusing  on the low-energy scattering, we perform DSEOM decompositions  on the imaginary part of the self-energy at the first fermionic Matsubara frequency  $\im\Sigma(k, i\omega_0)$. 
The DSEOM decompositions in the spin channel,  $\im\Sigma^{Q}_{sp}(\bk, i\omega_0)$ is displayed in Fig.~\ref{fig:fluc}  , as a function of  $Q=(\bq, i\Omega_n)$ for two typical dopings in the $T$-linear regime in  NFL, $p=0.18,0.2$. For comparison, the results at $p=0.08$ (PG) and $p=0.4$ (FL) are also shown. 
%The decompositions are carried out in both transfer momentum $\bq$ (Fig.5 A1-A4, B1-B4) and transfer frequency $i\Omega_n$  (Fig.5 C1-C4).
%The first column of Fig.5 (A1-A4) shows analysis on transform momentum $\bq$ at antinode [ $\bk=(0,\pi)$ ]. 
%We first note that although $\im\Sigma(\bk,i\omega_0)$ in NFL are much  weaker that in  PG ( Fig.4 ),
%the  DSEOM  decomposition $\im\Sigma^{\bq}_{sp}(\bk, i\omega_0)$  have surprisingly similar structure in $\bq$ space for NFL and PG :

We consider first the $p=0.18$ case. For both antinode
 [$\bk=(0,\pi)$, Fig.~\ref{fig:fluc} A2],  and node [$\bk=(\pi/2, \pi/2)$, Fig.~\ref{fig:fluc} B2], $-\im\Sigma^{\bq}_{sp}(\bk, i\omega_0)$  at different $\bq$ are extremely uneven.
The AFM wave vector $\bq=(\pi, \pi)$ component accounts for most of the low-energy scattering $-\im\Sigma(\bk, i\omega_0)$.  This means that in the NFL, most of the electronic scatterings are due to AFM fluctuations, since $\Sigma^{\bq}_{sp}(k) \propto \sum_{k^{\prime}} \langle S^{+}_{k}(-\bq) S^{-}_{k^{\prime}}(\bq) \rangle $. Moreover, from Fig.~\ref{fig:fluc} C2,  one learns from the  frequency decomposition that  the $\Omega = 0$ component dominates,
% components with higher energies ($\Omega_n >0$) account less than 10\% of the total weight.
suggesting the long-lived nature of the well-defined AFM fluctuations at this doping.

At a larger doping $p=0.20$,  the weight of the $\bq \neq (\pi, \pi)$ components grows, as shown in Fig.~\ref{fig:fluc}[A3, B3]. However the predominant role of the $\bq=(\pi, \pi)$ mode is not changed.
%We would like to point out that in the underdoped regime,  the opening of PG is  widely believed due to the strong AFM correlations by experiments and theories. From Fig. A3, B3, C3,  one can see that the DSEOM decompositions indeed  suggests AFM fluctuations dominates electronic scatterings.  This result is  surprising as one would consider AFM correlations are significantly reduced in the overdoped regime of Hubbard model.
In fact, we find that the  $\bq = (\pi, \pi)$ component always has the largest contribution to $-\im\Sigma^{\bq}_{sp}(\bk, i\omega_0)$ among different $\bq$ in the NFL,  even when $p$ is  further increased (see Supplementary Fig.~\ref{fig:fluc3}).
This result is somewhat surprising, as one would intuitively expect negligible AFM correlations in the overdoped regime. To clarify his problem, in Fig.~\ref{fig:sigma}b we plot the  spin-spin correlator $\langle S_{i+1}\cdot S_{i} \rangle$ between a pair of neighboring sites  $(i,i+1)$  as a function of doping $p$. This shows that, although  largely reduced by doping,   the strength of AFM correlations  remains significantly non-zero in the NFL. For example, at $p=0.2$,  $\langle S_{i+1}\cdot S_{i} \rangle \approx -0.06$, which is about $40\%$ of the value at $p=0.08$ in the PG.
Neutron scattering studies on LSCO  show that at $p=0.25$ in the NFL, the dynamical magnetic susceptibility still has fairly large intensity
at finite energy, whose magnitude is about half of that at $p=0.125$  in the PG~\cite{wakimoto}. Resonant inelastic scattering studies also reveal the persistence of spin excitations in the overdoped regime~\cite{LeTacon_Minola_Peets_2013}. This emphasizes again that the short-ranged AFM correlations should not be overlooked in the overdoped regime.

The decompositions for the PG and the FL are shown, respectively, in the first and last columns of Fig.~\ref{fig:fluc}.  In the PG,  $-\im\Sigma^{Q}_{sp}(\bk, i\omega_0)$ is similar to the NFL case, revealing again the importance of scattering off AFM fluctuations~\cite{gunnarsson2015,wu2017,taillefer2018}.
By contrast, in the FL phase,  a clear distinction between the 
NFL and PG cases is observed: $-\im\Sigma^{\bq / \Omega_n}_{sp}(k)$  with different $\bq / \Omega_n$ are more or less comparable. There is  no individual mode in $\bq / \Omega_n$  space that provides a dominant contribution to scattering.  This is expected, since  scattering in Fermi liquids should be seen as single-particle collisions rather  than  scattering off collective modes. Hence the two-particle \textit{spin} representation becomes \textit{inappropriate} to identify the source of  scattering in the FL.

We also performed DSEOM decompositions in other channels, and found  no  indication of any significant  \textit{charge} or  \textit{particle-particle} collective modes in the NFL (see Supplementary Sec.~\ref{sup:fluc} and Fig.~\ref{fig:fluc2}).  Therefore we conclude that in the NFL, most of the $T-$ linear electronic scattering comes from AFM fluctuations.

\section*{Discussion}
In recent ARPES measurements of Bi2212, it is found that the ARPES spectra near $p^{*}$ can be well fitted by a marginal Fermi liquid form for the self-energy $ -\im \Sigma (k,\omega) = T \phi (\omega /T) + b$~\cite{Chen1099}, which supports  our assumption of  $\omega/T$ scaling in the NFL state.
Moreover, we note that $\gamma^{\prime}_k(T)$ has similar slopes in $T$ at the node and at the antinode, which means that the inelastic part  ($T-$ dependent part) of the scattering rates,  $\gamma^{in}_k(T) \equiv \gamma^{\prime}_k(T)- \gamma^{\prime}_k(0)$ are isotropic in our study. For example at $p=0.18$, $\gamma^{in}_{N}(T)/\gamma^{in}_{AN}(T) \approx 1.1$, as shown in Fig.~\ref{fig:scatter}. This agrees with early ARPES results~\cite{Kaminski_2005} and very recent  angle-dependent magnetoresistance (ADMR) experiments on LSCO~\cite{grissonnanche2020}. We note that an immediate consequence of $\gamma_k(T)$ being perfectly linear-in-T in the NFL is that the dc resistivity $\rho_{T}$ without vertex corrections, can also have linear temperature dependence (see  Supplementary Sec.~\ref{Sup:resistivity}).

Where does the linear  $T-$  dependence come from? In the case of phonons,  when temperature $T$ is larger than about one third of the Debye frequency~\cite{hartnoll2021planckian}, The scattering rate increases like $T$ because  the number of bosonic scatterers grows linearly with $T$~\cite{Sadovskii_2021}. In the case of an antiferromagnetic QCP\cite{ millis1993, gegenwart2008,  lohneysen2007, Xu2020,dumitrescu2021}, the characteristic spin fluctuation frequency plays the role of the Debye frequency in the phonon case and it indeed vanishes. However it does not explain the $T-$linear scattering rate  in the case of weak interactions, since 
the  \textit{ electrons - spin fluctuations} scattering will be strong only at hot spots on the Fermi surface so that, barring disorder effects~\cite{Rosch:1999}, the resulting resistivity will be short-circuited by Fermi-liquid-like portions of the Fermi surface~\cite{Hlubina:1995}.

For the strong interaction, $U=7t$  that we considered, it can be  speculated that the lack of well-defined fermion quasiparticles leads to  spin fluctuations with overall vanishing characteristic frequency. Then, the argument that the number of scatterers scales like $T$ should hold. Since the magnetic correlation length is small in the over-doped regime~\cite{kastner1998},  the electrons on remains of the  Fermi surface can be all effectively scattered. Then the argument that the linear $T-$ dependence of the scattering rate is isotropic  on the Fermi surface will also hold.
In this case, dimensional analysis and Kanamori-Brückner screening suggest  (see Sec.~\ref{sup:dimensional} of  Supplementary)  that the coefficient of the linear $T-$ dependence of the scattering rate can be of order unity. But it does not need to be unity. In fact, we find a number about equal to three for the electron  scattering rate, and about ($1 \sim 2$) for quasiparticle scattering rate with the current parameters. So we call the strong-interaction case that we studied, a ``nearly Planckian liquid'' and we argue that Planckian dissipation likely not to be a fundamental limit to the inelastic electron scattering rate~\cite{Sadovskii_2021, Poniatowski}.

\section*{Conclusion}
To conclude, we  investigated the two-dimensional Hubbard model in the intermediate to strong interaction limit where a non-Fermi liquid phase is found to exist in the overdoped regime. We found that
the electronic scattering rate $\gamma_k(T)$  can have a perfectly linear  $T-$  dependence when doping $p$ is close to the pseudogap critical doping $p^{*}$. We also 
discovered that the antiferromagnetic fluctuations are responsible for the $T-$ linear electron scattering at low temperatures.
% and that this scattering can be larger than the so-called ``Planckian limit''. 

%\matmethods{

%Put methods in here.  If you are going to subsection it, use
%\verb|\subsection| commands.  Methods section should be less than
%800 words and if it is less than 200 words, it can be incorporated
%into the main text.
\section*{Method}
Our results for the two-dimensional Hubbard model are obtained using the dynamical
cluster approximation (DCA)~\cite{maier2005}, which is a cluster extension of the dynamical mean-field theory (DMFT). (See Supplementary Secs.~\ref{Sup:Geometry} to \ref{Sup:T*TFL} for details)  The DCA method captures short-ranged spatial 
correlations within the cluster exactly, while longer range spatial correlations are
taken into account by a dynamical mean-field, which can be represented by a momentum- and frequency- dependent Weiss 
field $g_0(\bk, i\omega_n)$. The effective cluster impurity problem starting from $g_0(\bk, i\omega_n)$ is solved by the Hirsch-Fye quantum Monte carlo method~\cite{hirsch1986monte}, 
which in general has a slightly better average sign as compared to the continuous-time quantum Monte Carlo method (CTQMC)~\cite{Gull:2011}.
Here we use  a discrete imaginary-time step $\Delta \tau = 0.071$. We have carefully verified  that this finite $\Delta \tau $ is small enough so that the Trotter errors  do not affect our result and conclusion, see Supplementary~Fig. \ref{fig:hfqmc}. Comparison with CTQMC result also  shows that our conclusion is not changed in the $\Delta \tau \rightarrow 0$ limit, see Supplementary~Fig. \ref{fig:ctqmc}.
 In this work, we typically use 60 DCA self-consistency iterations to get a converged Weiss field $g_0(\bk, i\omega_n)$,
or  equivalently a converged self-energy $\Sigma(\bk, i\omega_n)$. In the eight-site DCA approximation, the lattice self-energy is approximated by a patchwise-constant  self-energy $\Sigma(\bk, i\omega_n)$ in the Brillouin zone with eight different patches as shown in Supplementary~Fig. \ref{fig:geo}. Note that the antinodal and nodal regions are in distinct patches in this eight-site cluster scheme. We have verified that the $T-$ linear scattering rate also appears in  four-site DCA, and $4\times 4$ -site DCA calculations, namely, it can be checked explicitly for $T > 0.1$ that our results are insensitive to the cluster size, see Supplementary~Figs.~\ref{fig:gamma16}-\ref{fig:fluc16}.
%}

%\showmatmethods{} % Display the Materials and Methods section

\section*{Acknowledgment}
%\acknow{
We acknowledge discussions with Mathias Scheurer,  Andrey Chubukov, Nigel Hussey, Jake Ayres,  Antoine Georges,  Michel Ferrero, and Nils Wentzell.
 This work has been supported by  the funding from the National Natural Science Foundation of China (Grant No. 41030053), and  by the Natural Science Foundation of Guangdong Province (Grant No. 42030030), the Natural Sciences and Engineering Research Council of Canada (NSERC) under grant RGPIN-2019-05312 and by the Canada First Research Excellence Fund.   Part of the computational work was carried out at the National Supercomputer Center in Guangzhou (TianHe-2).
%}

%\showacknow{} % Display the acknowledgments section

\subsection*{References}
% Bibliography
\bibliography{nfl}

%\noindent\rule{\textwidth}{0.4pt}
%\bibliographystyle{pnas-new}
%\bibliography{nfl}
%\noindent\rule{\textwidth}{0.4pt}

\newcommand{\fref}[1]{Fig.~\ref{#1}}

\newcommand{\eref}[1]{Eq.~(\ref{#1})}

\newcommand{\sref}[1]{Sec.~(\ref{#1})}

\newcommand{\tref}[1]{Table~\ref{#1}}

\newcommand{\figcite}[1]{{\protect \cite{#1}}}

%%%%%%%%%% Merge with supplemental materials %%%%%%%%%%

%%%%%%%%%% Prefix a "S" to all equations, figures, tables and reset the counter %%%%%%%%%%

\newpage
\null
\newpage

\setcounter{equation}{0}
\setcounter{figure}{0}
\setcounter{table}{0}
\setcounter{page}{1}
\makeatletter

\renewcommand{\theequation}{S\arabic{equation}}
\renewcommand{\thefigure}{S\arabic{figure}}
\renewcommand{\thetable}{S\arabic{table}}
%\renewcommand{\bibnumfmt}[1]{[S#1]}
%\renewcommand{\citenumfont}[1]{S#1}
%%%%%%%%%% Prefix a "S" to all equations, figures, tables and reset the counter %%%%%%%%%%

\begin{widetext}

\begin{center}

    \textbf{\large Supplementary Materials: Non-Fermi liquid phase and linear-in-temperature scattering rate in overdoped   two dimensional Hubbard model}

\end{center}

\appendix

\author{W\'ei W\'u$^{1*}$, Xiang Wang$^{1}$ \& A.-M. S.Tremblay$^{2}$}

\section{Geometry of the DCA clusters} \label{Sup:Geometry}

The different DCA clusters that we have used are shown in the following Fig.~\ref{fig:geo}a shows the DCA patches in momentum space for the 8-site cluster.

\begin{figure}[h]
  \begin{center}
    \includegraphics[width=0.6\columnwidth]{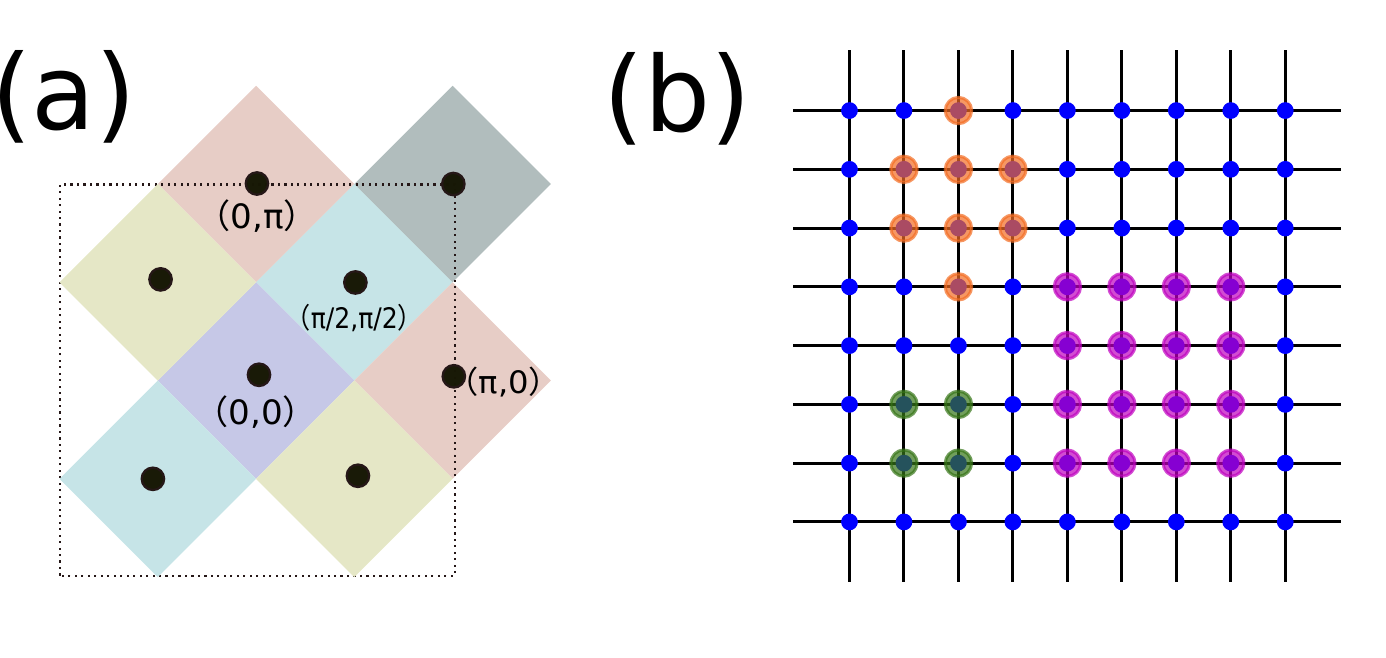}
  \end{center}
  \caption{
    \textbf{Geometries of the DCA clusters used in this work }
    \textbf{(a):} DCA patches in momentum space for the eight-site cluster.
    \textbf{(b):} Geometries of the $4-$, $8-$, and $16-$ site DCA clusters used in this work.}
\label{fig:geo}    
\end{figure}

\section{Analysis on the Trotter errors of the HFQMC solver}

In this work, we typically use $\Delta \tau = 0.071$ in the Hirsch-Fye impurity solver~\cite{hirsch1986monte}. We have carefully verified  that this finite $\Delta \tau $ is small enough  that the Trotter errors  do not affect our result and conclusion, as shown in \fref{fig:hfqmc}. Comparison with CTQMC results also  shows that our conclusion is not changed in the $\Delta \tau \rightarrow 0$ limit,  as shown in  \fref{fig:ctqmc}.

\begin{figure}[h]
  \begin{center}
    \includegraphics[width=\columnwidth]{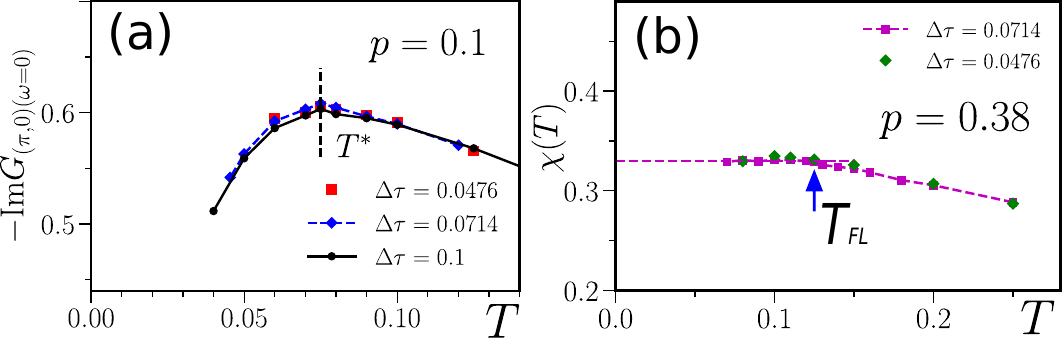}
  \end{center}
  \caption{
    \textbf{The pseudogap temperature $T^{*}$ and Fermi liquid temperature $T_{FL}$ are converged in HFQMC at $\Delta \tau = 0.0714$.}
    \textbf{(a):} The imaginary part of the antinodal Green's function $-\im G_{(\pi, 0)}(\omega=0)$ shows a maximum as temperature $T$ decreases 
    at the same $T$ for $\Delta \tau$ = 0.1 (circle), 0.0714 (diamond), and 0.0476 (square), suggesting our  $T^{*}$ data in the main text at $\Delta \tau = 0.0714$ is converged.
    \textbf{(b):} The paramagnetic susceptibility $\chi(T)$ saturates at the same $T$ for $\Delta \tau$ = 0.0714 (square), and 0.0476 (diamond), suggesting that our $T_{FL}$ data at $\Delta \tau = 0.0714$ is also converged. Here $U=7t, t^{\prime} =-0.2t$.}
\label{fig:hfqmc}    
\end{figure}

\begin{figure}

  \begin{center}
    \includegraphics[width=\columnwidth]{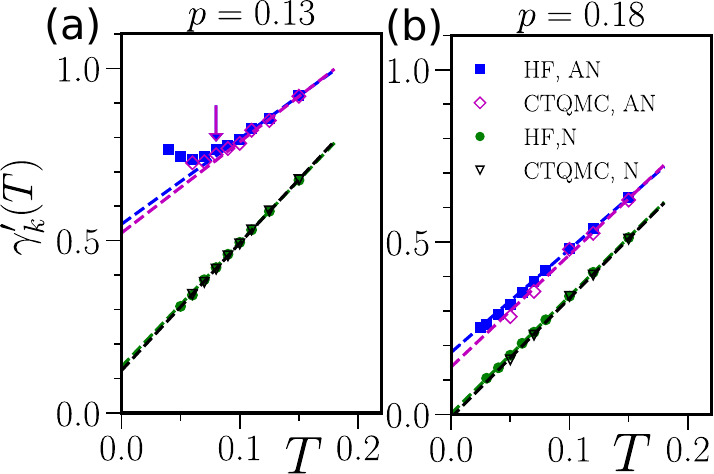}
  \end{center}
 \caption{ 
    \textbf{Comparison of the HFQMC and CTQMC results for the electronic scattering rate $\gamma^{\prime}_{\bk}(T)$ . }
    \textbf{(a):} At doping level $p=0.13$ in the PG.
    \textbf{(b):}  At doping level $p=0.18$ in the NFL. For the node, the differences in $\gamma^{\prime}_{\bk}(T)$ between CTQMC  and HFQMC are negligible at both dopings.
     For the antinode, $\gamma^{\prime}_{\bk}(T)$ of CTQMC  is slightly smaller than that of HFQMC at low temperatures. 
     In the PG ($p=0.13$), the temperatures $T^{*}_{o}$ where  $\gamma^{\prime}_{\bk}(T)$ starts to  deviate from  linearity (marked by arrows)  are essentially the same for CTQMC and HFQMC. 
     In the NFL ($p=0.18$), CTQMC and HFQMC results both show $T-$ linear  antinodal $\gamma^{\prime}_{\bk}(T)$, despite the slightly larger slope of the antinodal  $\gamma^{\prime}_{\bk}(T)$  of CTQMC. Hence our conclusion drawn from $\Delta \tau = 0.0714$ result that for $ 0.17 \lesssim p \lesssim 0.20$, the electronic scattering rate is linear in $T$ is  not changed when  $\Delta \tau \rightarrow 0$ is extrapolated. Here $U=7t, t^{\prime} =-0.2t$.}

\label{fig:ctqmc}    
\end{figure}

\section{Pseudogap temperature $T^{*}$ and Fermi liquid temperature $T_{FL}$}\label{Sup:T*TFL}
In this work the pseudogap $T^{*}$ is identified as the temperature  where
the antinodal zero- frequency spectral function ( obtained by extrapolation~\cite{wu2018} ) displays a maximum. Thus below $T^{*}$, the antinodal spectral intensity decreases, denoting the opening of a pseudogap. $T_{FL}$ is defined as the temperature where the paramagnetic susceptibility $\chi(T)$ ( Knight shift ) saturates while decreasing temperature $T$, as shown in ~\fref{fig:tt}.

\begin{figure}
  \begin{center}
    \includegraphics[width=\columnwidth]{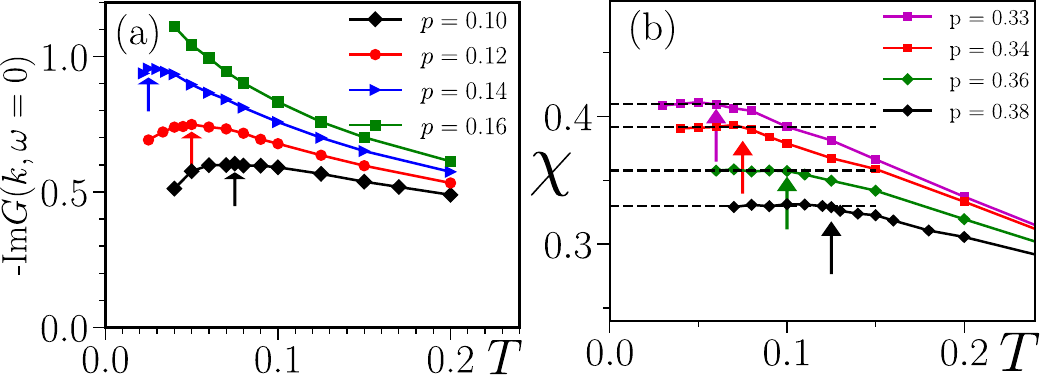}
  \end{center}
  \caption{
    \textbf{Pseudogap temperature $T^{*}$ and Fermi liquid temperature $T_{FL}$ at $U=7t, t^{\prime} = -0.2t$ for various dopings }
    \textbf{(a):} Pseudogap temperature $T^{*}$ is defined as the temperature where $- \im G(k, \omega=0)$ at $k=(0,\pi)$ reaches  a maximum. Here $- \im G(k, \omega=0)$ 
    is obtained by linear fitting  in Mastsubara frequency space using Green's functions at the  lowest two frequencies. 
    \textbf{(b):}  Fermi liquid temperature $T_{FL}$ is defined as the  temperature where the paramagnetic susceptibility $\chi(T) = \frac{1}{N} \sum_{i,j} \int_{0}^{\beta} \langle S_{i}( \tau ) \cdot S_{j}(0) \rangle  d \tau$ saturates when decreasing $T$. The dashed lines denote the average  $\chi(T)$  values of the last four $T$ points for each doping.
     Here the numerical data of $\chi(T)$ is actually obtained by computing the paramagnetic  response  to  a small uniform magnetic field to the system.  The inducing magnetic field is chosen small enough to ensure a linear response.  Arrows in the two subplots indicate
     $T^{*}$  (Left) and $ T_{FL}$ (Right) respectively.
 }
\label{fig:tt}    
\end{figure}

\section{$\omega/T$ scaling of the self-energy} \label{sup:omega}
In Fermi liquids the imaginary part of the self-energy obeys  $\omega/T$ scaling. It can occur more generally in strongly correlated systems that physical quantities display $\omega/T$ scaling when  the relevant characteristic energy scales vanishes~\cite{schroder2000onset, sachdev_book, parcollet1999,gegenwart2008}. We find that 
in the NFL regime of the overdoped Hubbard model,  numerical data  suggests $\omega/T$ scaling behavior of the imaginary part of the self-energy $\im\Sigma(\bk,\omega)$ in real-frequency space, as discussed below.

It has been shown that   $\omega/T$ scaling of $\Sigma^{''}(\bk,\omega)\equiv \im\Sigma(\bk,\omega)$ at low-energies leads to 
$\tau/\beta$ scaling~\cite{dumitrescu2021} near $\tau \sim 0.5\beta$ when translated in imaginary time $\Sigma(\bk,\tau)$. This is because,
\begin{eqnarray}
\Sigma(\bk, \tau) =  \int \Sigma^{''}(k,\omega)K_{\beta}(\omega, \tau)\frac{d\omega}{\pi} \\ 
= \int \frac{\Sigma^{''}(k,\omega)e^{-\tau \omega}}{1+e^{\beta\omega}}\frac{d\omega}{\pi}\
 =  \frac{T}{\pi}\int \frac{\Sigma^{''}(k,\omega)e^{-\frac{\tau}{\beta} \times( \omega/T)}}{1+e^{\omega/T}}d\frac{\omega}{T}
\end{eqnarray}
\textit{i.e.}, the integral kernel $ K_{\beta}(\omega, \tau)$ can be rewritten as a function of $\tau/ \beta$ and $\omega/T$. Therefore when the $\omega-$ dependence of  $\Sigma^{''}(\bk,\omega)$ can be expressed as a function of   $\omega/T$, $\Sigma(\bk, \tau)$ will follow $\tau/\beta$ scaling. Note that when $\tau \sim 0.5\beta$, the integral kernel $ K_{\beta}(\omega, \tau)=e^{-\tau \omega}/(1+e^{\beta\omega})$ is a bell-shaped function in $\omega$, which  essentially  collects the  low-energy weight of $ \Sigma^{''}(\bk,\omega)$ between  $ -4T \lesssim \omega \lesssim 4T$. 

Here we use a slightly different method to show the $\omega/T$ scaling behavior of $\Sigma^{''}(\bk,\omega)$ from $\Sigma(\bk,\omega_n)$ data in Matsubara frequencies. Fitting $\im\Sigma(\bk, i\omega_n)$ data to the first three Mastsubara frequencies,  $\im\Sigma(\bk, i\omega_0)$, $\im\Sigma(\bk, i\omega_1)$, $\im\Sigma(\bk, i\omega_2)$ to a quadratic function of $\omega_n$, and then extrapolating to small frequencies $\omega_m$ ( $\omega_m < \omega_0 = \pi T$ ), we obtain an extrapolated self-energy $\im \Sigma^{(2)}(\bk, i\omega_m )$ that is equal to,
\begin{eqnarray}
\im \Sigma^{(2)}(\bk, i\omega_m ) =  \int \Sigma^{''}(k,\omega )K_{T}(\omega ,\omega_m)\frac{d\omega}{\pi}  \label{eq:extrap}
\end{eqnarray}
with the integral kernel $ K_{T}(\omega, \omega_m)$ given by,
\begin{eqnarray}
K_{T}(\omega, \omega_m) = A(\omega,T) \omega_m^2 + B(\omega, T)\omega_m + C(\omega ,T)\nonumber \\ 
A(\omega ,T) =\im [ \frac{1}{8\pi^2 T^2(\omega -i\pi T)}-\frac{1}{4\pi^2T^2(\omega -3i\pi T)}+\frac{1}{8\pi^2 T^2(\omega -5i \pi T)} ] \nonumber \\
B(\omega ,T) =\im [  -\frac{1}{\pi T(\omega -i\pi T)}+\frac{3}{2\pi T(\omega -3i\pi T)}-\frac{1}{2\pi T(\omega -5i \pi T)} ] \nonumber \\
C(\omega ,T) = \im [ \frac{15}{8(\omega -i\pi T)}-\frac{5}{4(\omega -3i\pi T)}+\frac{3}{8(\omega -5i \pi T)} ]. \label{eq:second}
\end{eqnarray}
The kernel $ K_{T}(\omega, \omega_m)$  is also a bell-shaped function in energy $\omega$ whose weight is mainly between  $ -4T \lesssim \omega \lesssim 4T$,  when $\omega_m $ are small ( $|\omega_m|<\pi T$, see Fig.~\ref{fig:ken}). Note that $ K_{T}(\omega, \omega_m)$ can be rewritten as a  function of  $\omega/T $ and $\omega_m/T$,
 $$ K_{T}(\omega, \omega_m)=\kappa(\omega/T, \omega_m/T).$$
  Thus $\im \Sigma^{(2)}(\bk, i\omega_m )$  obtained from the integral of equation (\ref{eq:extrap}) exhibits   $\omega_m /T$ scaling  at small $\omega_m$,   if  $\Sigma^{''}(\bk,\omega)$ has  $\omega/T$ scaling  at low-energies .

According to the above analysis, one can therefore extrapolate $\Sigma(\bk, i\omega_n)$  using a second order polynomial fit in Matsubara frequency space to obtain $\im \Sigma^{(2)}(\bk, i\omega_m )$ at small $\omega_m$  and verify whether $\Sigma^{''}(\bk,\omega)$ obeys  $\omega/T$ scaling at low-energies. Our DCA result is shown in Fig.~\ref{fig:scale} where one can see that for $p=0.18$ in the NFL, $\im \Sigma^{(2)}(\bk, i\omega_m )$ \big(normalized by $\im \Sigma^{(2)}(\bk, i\omega_m=0)$\big) at different temperatures $T$ indeed collapses nicely to a single scaling function of $\omega_m/T$ . In other words, $\im \Sigma^{(2)}(\bk, i\omega_m )/\im \Sigma^{(2)}(\bk, i\omega_m = 0) = S(\omega_m/T)$ holds at different $T$ for small $\omega_m$, where $S(\omega_m/T)$ appears to be essentially a linear function of $\omega_m/T$ according to Fig.~\ref{fig:scale}.  This result unambiguously shows that in the NFL,  $\Sigma^{''}(\bk,\omega)$ does follow  $\omega/T$ scaling at low-energies.

\begin{figure}
  \begin{center}
    \includegraphics[width=0.6\columnwidth]{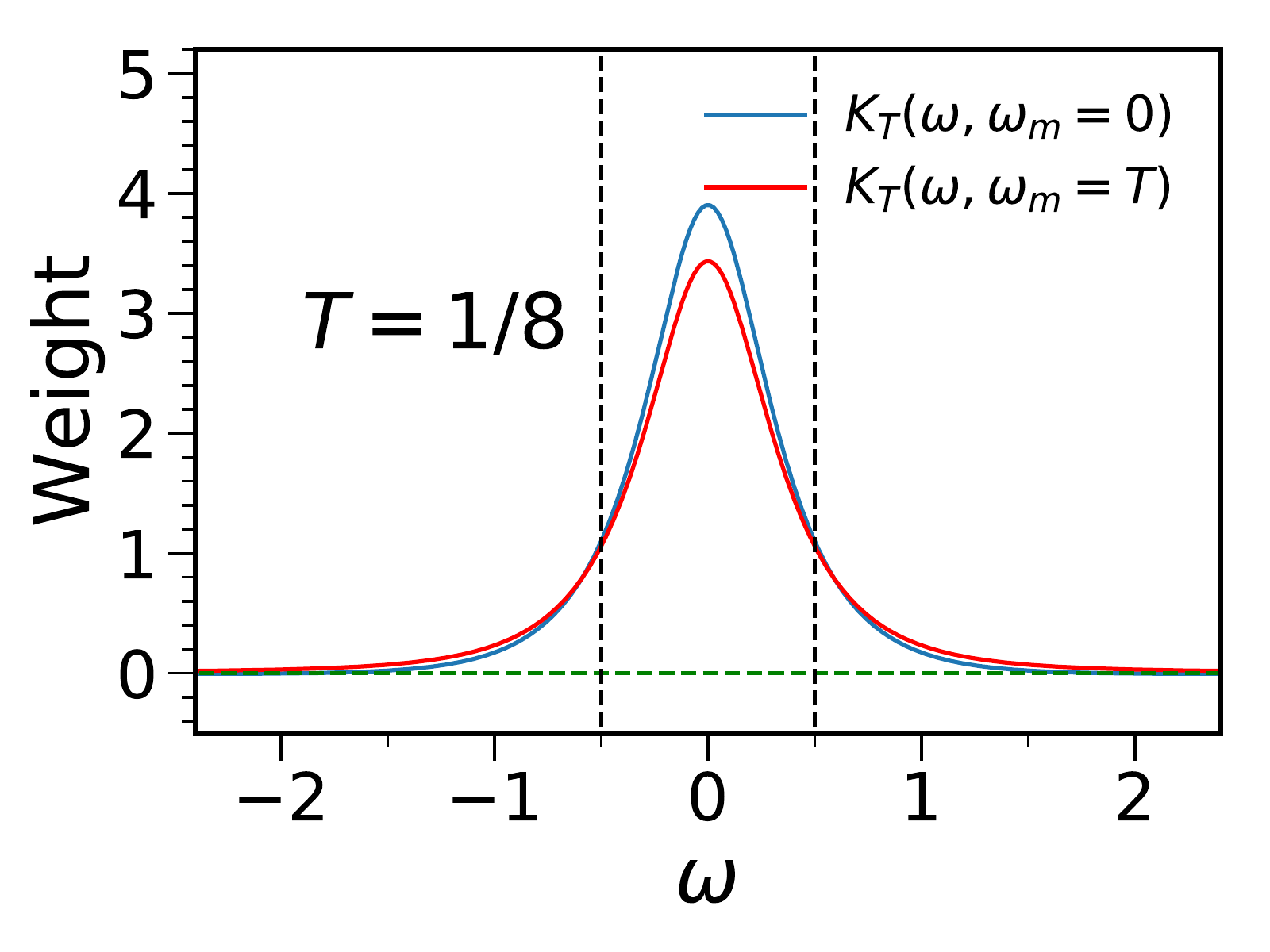}
  \end{center}
  \caption{
    \textbf{The  $ K_{T}(\omega, \omega_m)$  function at two different  $\omega_m$.} 
    Vertical dashed lines show $\omega=-4T$ and  $\omega=4T$ respectively.
   }

\label{fig:ken}    
\end{figure}

\begin{figure}
  \begin{center}
    \includegraphics[width=\columnwidth]{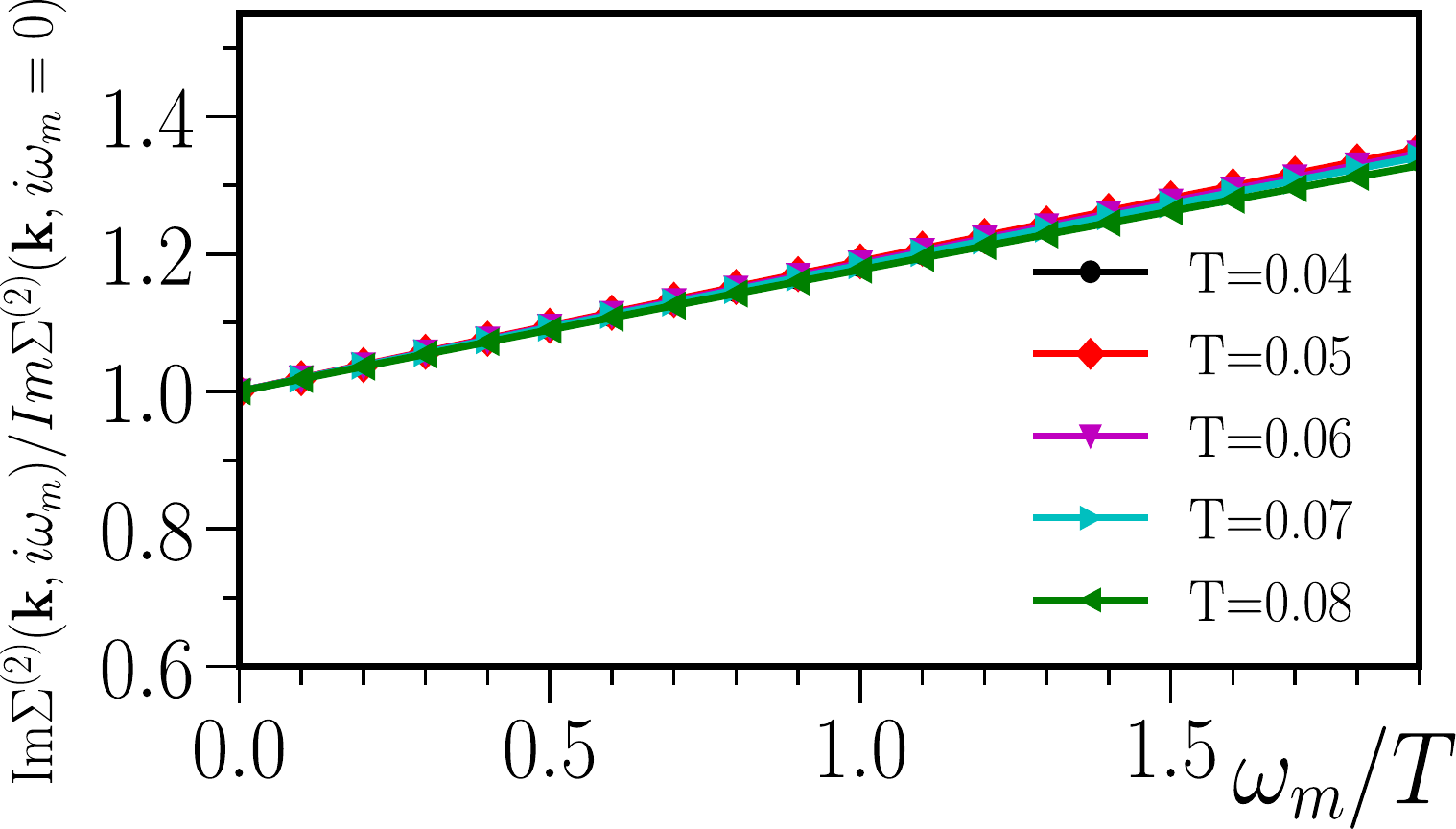}
  \end{center}
  \caption{
    \textbf{$\omega_m/T$ scaling of the imaginary part of the extrapolated self-energy $\im\Sigma^{(2)}(\bk,i\omega_m)$ at small imaginary frequencies $\omega_m$ .  
   }
    $\im\Sigma^{(2)}(\bk,i\omega_m)$ is obtained from a second order  polynomial extrapolation of   $\im\Sigma(\bk,i\omega_n)$ in Matsubara frequencies. 
    Here $u=7t, t^{\prime} = -0.2t , p=0.18$ and $\bk=(\pi, 0)$.
   }
\label{fig:scale}    
\end{figure}

We have shown above that 
the energy dependence of  $\im \Sigma(k,\omega)$ follows $\omega/T$ scaling at low-energies in the NFL. If we assume that   $\Sigma^{''}(k,\omega) \equiv \im \Sigma(k,\omega) $ can be written as $\Sigma^{''}(k,\omega)=-T^{\alpha}\phi(\omega/T)-b$, where $\phi(\omega/T)$ is an unknown analytic function of $\omega/T$, and $b$ is a constant, it is straightforward to prove that $\im \Sigma^{(n)}(k, \omega=0)$ extrapolated with an $n-$ order  polynomial fit of  $\im \Sigma(k, i\omega_n)$  in Matsubara frequency space will have the same $T-$ dependence as the true scattering rate $-\im \Sigma(k, \omega=0)$. To be specific,  take the second order extrapolation 
$$\gamma^{\prime}_k \equiv -\mathrm{Im}\Sigma^{(2)}(k, i\omega_m=0) =-\im [1.875\Sigma(k,   i\omega_0)-1.25 \Sigma(k, i\omega_1)+ 0.375 \Sigma(k,  i\omega_2)], $$ 
then the integral kernel $K_T(\omega,\omega_m=0)$  reads,
 % To demonstrate this point, we note that a second-order polynomial extrapolation of selfenergy  $-\mathrm{Im}\Sigma^{(2)}(k, \omega=0)=-\im [1.875\Sigma(k,   i\omega_0)-1.25 \Sigma(k, i\omega_1)+ 0.375 \Sigma(k,  i\omega_2)]  \doteq \beta\Sigma(k, \tau = 0.5\beta)$ reads,
\begin{eqnarray}
K_T(\omega ,0) =C(\omega ,T) = \sum_{n=1}^{n=3} \frac{c_n \omega_n}{\omega ^2 + {\omega_n}^2}    & c_{1,2,3} = [1.875, -1.25, 0.375] 
\end{eqnarray}
(See also Eq.~\ref{eq:second} and Fig~\ref{fig:ken}). Hence, as long as $\omega/T$ scaling applies in the
range  $|\omega | < \omega_M \approx 4T$ for $\Sigma^{''}(k,\omega)$, we have for $ \gamma^{\prime}_k$
\begin{eqnarray}
\gamma_{k}^{\prime}=\frac{1}{\pi}\int_{-\omega_{M}}^{\omega_{M}}\sum_{n=1}^{n=3}\frac{c_{n}\omega_{n}}{\omega^{2}+{\omega_{n}}^{2}}[T^{\alpha}\phi(\frac{\omega}{T})+b]d\omega\\
=\frac{T^{\alpha}}{\pi}\int_{-\omega_{M}/T}^{\omega_{M}/T}\sum_{n=1}^{n=3}\frac{c_{n}(2n+1)\phi(\frac{\omega}{T})}{(\frac{\omega}{T})^{2}+(2n+1)^{2}\pi^{2}}d(\frac{\omega}{T})+b\\
=aT^{\alpha}+b
\end{eqnarray}
where the above integral over $\omega/T$ yields a constant $a \pi $ because the integrand is a function of $\omega/T$. Comparing to the true electronic scattering rate $\gamma_k \equiv -\Sigma^{''}(k,\omega =0 ) = \phi(0)T^{\alpha}+b$, we see that $\gamma^{\prime}_k  =  aT^{\alpha}+b$ indeed captures correctly the $T-$ dependence of $\gamma_k $. 

Note that if $\Sigma^{''}(k,\omega)$ is constant over the frequency range $|\omega| \lesssim \omega_M \approx 4T$, or namely, if $\phi(\omega/T)$ becomes $\omega -$ independent in $|\omega| \lesssim \omega_M$,  the above integral will lead to $\phi(0) \approx a$. Thus $\gamma^{\prime}_k  \approx \gamma_k $ in such situation. For the marginal Fermi liquid selfenergy~\cite{mfl}, $\im \Sigma(\bk,\omega)= \alpha \mathrm{max}(|\omega|, \pi T)$ which becomes $\omega-$ dependent when $|\omega|  > \pi T$.  Therefore in general $\phi(0) < a$ can be speculated for marginal Fermi liquid,  and $\gamma^{\prime}_k$ should have a slope in $T$ slightly larger than that of the true electron scattering rate $\gamma_k$.

%\end{align}
%while the true scattering rate is $\gamma_k = \pi \phi(0)T^{\alpha}+b$. Namely, one can find the \textit{ exact} exponent $\alpha$  of the $T-$ dependence of $\gamma_k $ from $ \gamma^{\prime}_k $ , despite the constant coefficients,  $\phi(0)$ and $a$ are not known. See supplementary for more  justifications of the   $\omega/T$ scaling  from both theory and numerical data.  

\section{$T-$ linearity of the scattering rate in the NFL phase. }\label{sup:tlinear}

In the NFL, the electronic scattering rate $\gamma^{\prime}_k(T)$ can in general display a linear temperature dependence $\gamma^{\prime}_k(T) =aT+b$ .
However, in the underdoped cases the temperature where $\gamma^{\prime}_k(T)$ starts to deviate from linearity (marked as $T^{*}_o$ in Fig.~\ref{fig:scatter}a-b) is higher than $T^{*}$. This means that when $p$ just surpasses $p^{*}$, $\gamma^{\prime}_k(T)$ can still can deviate from linearity  (since $T^{*}_o$ is finite), although the PG temperature $T^{*}$ vanishes. Extrapolating $T^{*}_o$ to zero, as shown in \fref{fig:allanti},  we find that the minimal doping where $\gamma^{\prime}_k(T) $ can preserve  $T-$ linearity  in the $T=0$ limit is around $p_{L} \simeq 0.17$, which is slightly larger than $p^{*}=0.16$ where the pseudogap ends. 
  
Note that in experiments, there are usually different ways to define $T^{*}$. For example, sometimes $T^{*}$ is defined as  the temperature where the dc resistivity $ \rho (T)$ departs from linearity~\cite{taillefer2018}. This effectively defines $T^{*}_o$ as the pseudogap temperature, which would lead to a slightly different $p^{*} \equiv p_{L} = 0.17$. 

On the overdoped side of the NFL, we find that when $p>0.2$,    $\gamma^{\prime}_k(T) $ can also deviate from linear-in-$T$ behavior  at very small $T$ owing to the  onset of Fermi liquid physics, even though there is no finite $T_{FL}$.   As shown in Fig.~\ref{fig:scatter}f for $p=0.28$, extrapolating   $\gamma^{\prime}_k(T) $ to zero $T$ using $\gamma^{\prime}_k(T)=aT+b $  leads to a nonphysical $\gamma^{\prime}_k(T=0) <0 $,  signaling the failure of a purely linear function to describe $\gamma^{\prime}_k(T)$ in the $T\rightarrow 0 $ limit. Hence  for $p>0.2$, higher order corrections, such as quadratic or cubic  terms could  develop in $\gamma^{\prime}_k(T)$   at  small $T$,  as a result of Fermi liquid onset. 
To summarize the above analysis, we find that the scattering rate $\gamma^{\prime}_k(T)$  in the NFL phase  displays perfect $T-$ linear behavior as $T\rightarrow 0$  in the doping range of $ 0.17 \lesssim p \lesssim 0.20$ for  $t^{\prime}/t = -0.2$.

\begin{figure}
  \begin{center}
    \includegraphics[width=\columnwidth]{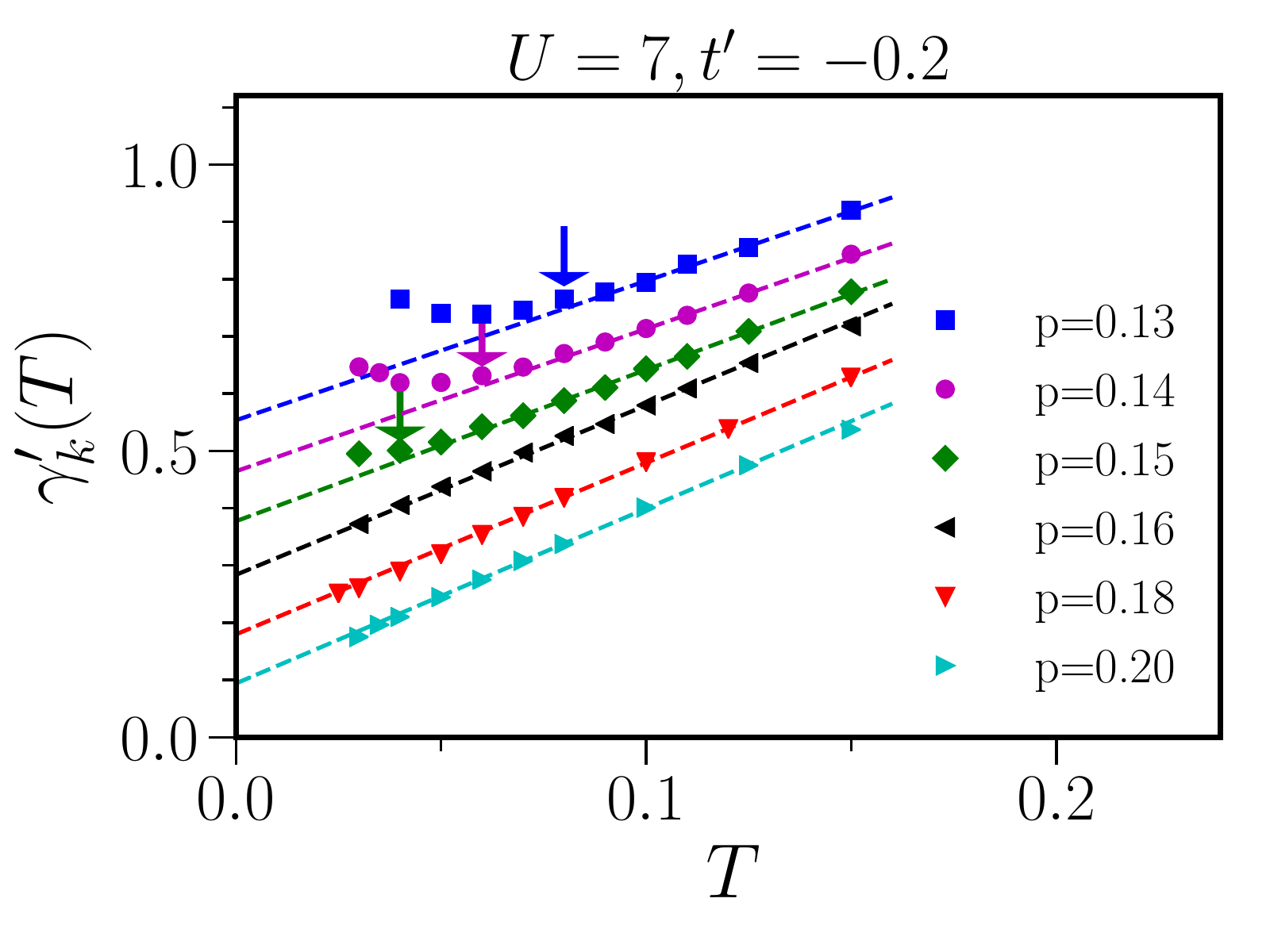}
  \end{center}
  \caption{
    \textbf{ Antinodal scattering rate $\gamma^{\prime}_k(T)$ [$\bk=(0,\pi)$] as a function of temperature $T$ for various dopings.   }
      Arrows indicate the temperature  $T^{*}_o$,  where $\gamma^{\prime}_k(T)$ starts to deviate from linearity. Extrapolating $T^{*}_o$ to zero, we estimate that the minimal doping where  
       $\gamma^{\prime}_k(T)$ is linear in the $T \rightarrow 0$ limit is $p \approx 0.17$. Dashed lines show linear fits (least square)  of $\gamma^{\prime}_k(T)$
       in the temperature range $T= (0.09-0.125)$.
    }
\label{fig:allanti}    
\end{figure}

\section{Quasiparticle scattering rate} \label{sup:quasi}
In the main text, we have investigated the electronic scattering rate $ \gamma_k \equiv -\im\Sigma(\bk, \omega=0)$. To study the
the quasiparticle scattering rate or inverse  quasiparticle life-time  $1/\tau_k = z_k \gamma_k$
, one needs  to also find out the quasiparticle weight $z_k $. To obtain $z_k $, here we  assume that the Green's functions at $k_F$ have a quasiparticle form as $G(\bk_F,\omega) = z_k/[\omega -iz_k \im \Sigma(\bk_F,\omega)]$ at low-energies, and  the imaginary part of the selfenergy at low-energies is assumed to be of the marginal Fermi liquid (MFL)  type,  $-\im \Sigma(\bk,\omega)= \alpha \mathrm{max}(|\omega|, \pi T)+b$~\cite{Varma2020}. With this hypothesis,  we fit the Green's function data in imaginary time space  $G(\bk_F,\tau)$ by,
\begin{eqnarray}
G(\bk_F,\tau) =  \int d \omega \frac{A(\bk_F,\omega)e^{-\omega\tau}}{1+e^{-\omega \beta}} \nonumber \\
  A(\bk_F,\omega)  =   -\frac{1}{\pi} \frac{\alpha \mathrm{max}(|\omega|, \pi T)+b}{(\omega /z_k )^2+[\alpha \mathrm{max}(|\omega|, \pi T)+b]^2} \label{eq:fit}
\end{eqnarray}
in vicinity the of $\tau \sim 0.5\beta$  [to filter out the low-energy behaviors of $G(\bk_F,\omega)$] and to find out the optimal free parameters $z_k$ and $\alpha$. The value of the constant $b$ is fixed as the extrapolated value of $\gamma^{\prime}_{\bk}(T)$ in the $T \rightarrow 0$ limit   from Fig.~\ref{fig:scatter}. Therefore the quasiparticle scattering rate $1/\tau_k$ can be identified as  $1/\tau_k = \alpha z_k \pi T+bz_k $, as shown in Fig.~\ref{fig:fit}. \
   Fig.~\ref{fig:life} shows  $1/\tau_k$ as a function of $T$ in the $T-$ linear regime, \textit{i.e.,} $p=0.18$ and $p=0.20$.  One can clearly see that at these dopings, $1/\tau_k \simeq CT$, with $C \sim (1\sim2)$,  namely the inverse quasiparticle lifetime is proportional to absolute temperature $T$ with a coefficient  $C$ close to unit. We would like to stress that here 
 the value of $C$ is    apparently dependent on the  doping level, and  is different between node and antinode.
   
 Performing numerical analytic continuations \big(such as the maximum entropy method (MEM)\big) on the Green's functions $G(\bk, i\omega_n)$,  one can
 obtain the spectral functions $A(\bk, \omega)$, and  identify the quasiparticle scattering rate $1/\tau_k$  as the half width at half maximum (HWHM) of the low-energy peak of $A(\bk, \omega)$. 
Fig.~\ref{fig:mem} shows MEM result~\cite{omaxent} on  the $T-$ dependence of $1/\tau_k$ , which suggests $1/\tau_k \sim 2.5T$ for antinode at $p=0.18$. This is in good agreement with the  result from fitting the Green's function $G(\bk_F,\tau)$, which suggests $1/\tau_k \sim 2.2T$ for antinode at $p=0.18$  in Fig.~\ref{fig:life}.

\begin{figure}
  \begin{center}
    \includegraphics[width=\columnwidth]{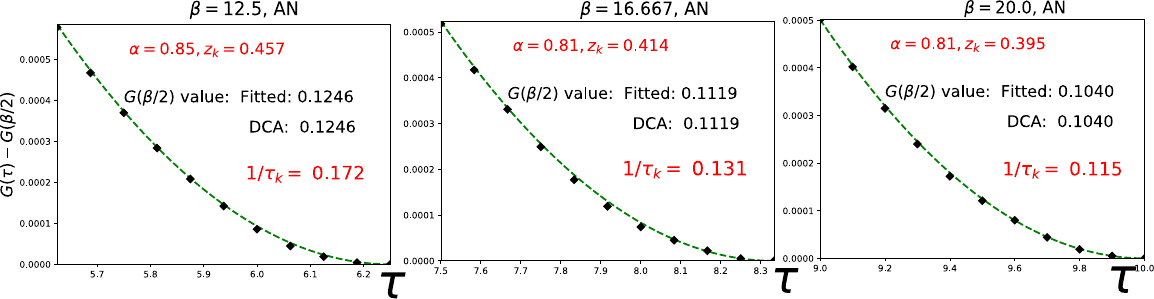}
  \end{center}
  \caption{
    \textbf{Fitting the Green's function $G(\bk, \tau)$ near $\tau \sim 0.5\beta$ with a MFL type of self-energy in theNFL.}
    Curves show  fitted $G(\bk_F,\tau)$ using optimized $\alpha$ and $z_k$ \big(shifted by $G(\bk_F,\tau=0.5\beta) $, see Eq.~\ref{eq:fit}\big), 
    while symbols are DCA data points.
   From left to right: $T=0.08, 0.06,0.05$. Here $U=7t, t^{\prime} = -0.2t, p=0.18$ and $\bk_F \approx (\pi,0.20)$, in the antinodal (AN) direction. The optimized 
   free parameters $\alpha$, $z_k$, and quasiparticle scattering rate $1/\tau_k $ are labeled in corresponding subplots. See also Fig.~\ref{fig:life}.
   } \label{fig:fit}
\end{figure}

\begin{figure}
  \begin{center}
    \includegraphics[width=\columnwidth]{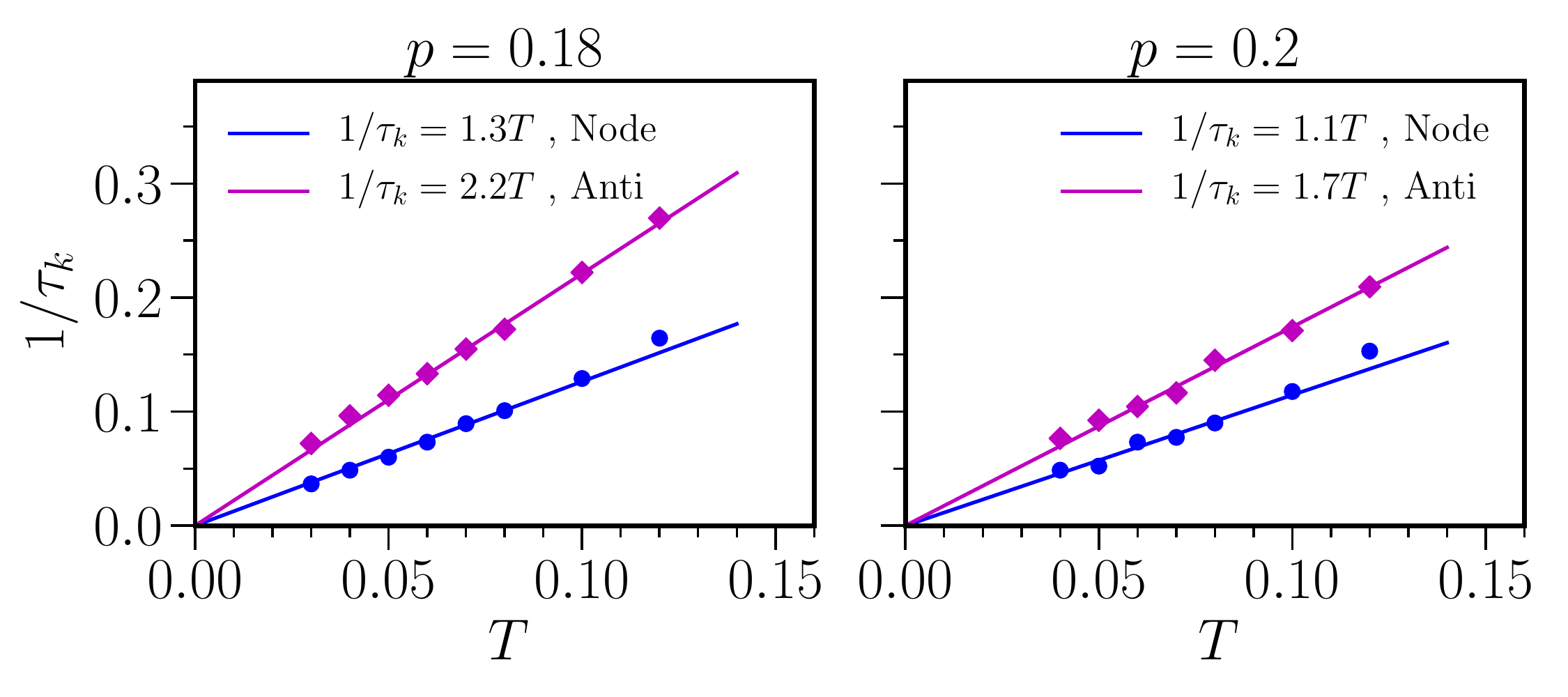}
  \end{center}
  \caption{
    \textbf{Quasiparticle scattering rate $1/\tau_k = -z_k \im\Sigma(k, \omega=0)$ as a function of temperature $T$ in NFL. }  Here $U=7t, t^{\prime} = -0.2t, p=0.18$.
    \textbf{(a):} For $p=0.18$  \textbf{(b):}  For $p=0.20$. $1/\tau_k$ is obtained by fitting the Green's function  $G(\mathbf{k}_F,\tau)$, see Fig.~\ref{fig:fit}.
   } \label{fig:life}
\end{figure}

\begin{figure}
  \begin{center}
    \includegraphics[width=\columnwidth]{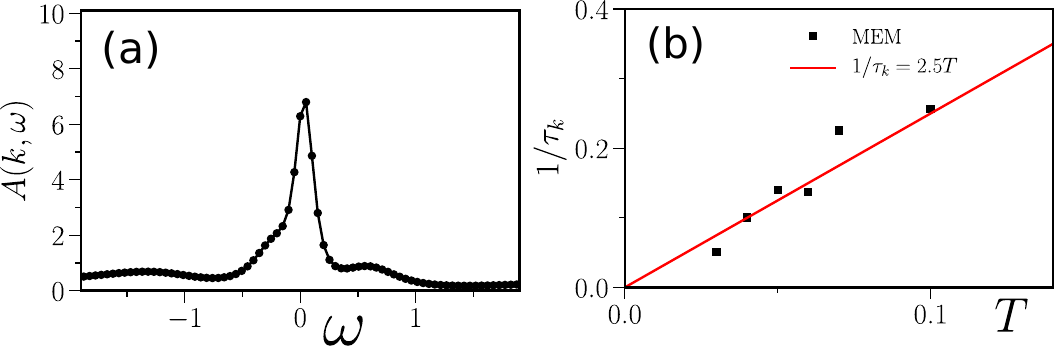}
  \end{center}
  \caption{
    \textbf{Spectral function $A(k,\omega)$ and quasiparticle scattering rate $1/\tau_k $ as a function of temperature $T$ by maximum entropy method analysis (MEM) in the NFL.} 
    \textbf{(a):} Spectral function $A(\bk,\omega)$ at $p=0.18, T=0.06$  \textbf{(b):}  Quasiparticle scattering rate $1/\tau_k $ in the antinodal direction.  Here $U=7t, t^{\prime} = -0.2t, p=0.18$, $\bk \approx (\pi,0.20)$.
   } \label{fig:mem}
\end{figure}

\section{Temperature dependence of the dc resistivity }\label{Sup:resistivity}
In this work, we have concentrated on the single-particle properties of the doped Hubbard model. 
% Here we briefly discuss 
% the temperature dependence of dc resistivity in the NFL state. 
The dc conductivity without vertex correction can be written as,
\begin{equation}
\sigma_{xx} = -2\pi \sum_{\bk}\big( \frac{\partial \epsilon_k}{\partial k_x} \big)^2 \int d\omega \frac{\partial f(\omega)}{\partial \omega} A^2(\bk,\omega)=\sum_{\bk} \sigma_\bk. \label{eq:conduct}
\end{equation}
Thus, the dc conductivity can be interpreted as the series addition of conductivities $\sigma_\bk$ (parallel addition of resistivities) defined for each value of wave vector $\bk$. A rigorous calculation of the conductivities $\sigma_\bk$ requires the inclusion of vertex corrections. Assuming that vertex corrections do not modify the linear dependence of the scattering rate that we found, this implies that  $R_\bk=1/\sigma_\bk\sim\gamma_\bk\sim aT+b$. In this work, we found that the coefficient $a$ is in general the same for antinode and node. If the  $T-$ independent scattering rate $b$ is also isotropic on different $\bk$, the total dc resistivity
$R = 1/ ( \sum_{\bk} 1/R_{\bk} )$ will be simply linear in temperature, considering the effective band dispersion does not change with $T$ (see~Fig.\ref{fig:ek}).  However,  in the antinodal direction, we extrapolated a finite intercept $b$ different from that of the node. Therefore if $b$ is finite or goes to infinity at $T=0$, the asymptotic behavior of the resistivity remains linear at low temperature with a crossover to another linear regime at high temperature~\cite{cooper2009}. If $b$ vanishes as in a Fermi liquid, the asymptotic behavior recovers the Fermi liquid form, unless the linear component also remain, in which case the resistivity is, again, asymptotically linear at low temperature.

It is worth noting that  ADMR experiments in Nd-LSCO~\cite{grissonnanche2020} have  also found  that the inelastic part ( $T-$ independent part)  of the antinodal scattering rate differs  from the  nodal one (in the temperature regime $T<30$K where the dc resistivity is perfectly linear in $T$). Given the uncertainties with vertex corrections, in this work we focus on the  scattering rate and leave the relation between the single-particle scattering rate and the transport properties for future study.

% Evaluating above integral for an isotropic system leads to the relation for the dc resistivity without vertex corrections $\rho_{T}=-6\im \Sigma(k_F, 0)/e^2 {v_F}^2 N(0)$~\cite{Varma2020,sadovskii2020}, where $v_F$ and $N(0)$ are respectively the bare Fermi velocity and bare density of states on Fermi level. Here we have antinode and node in different DCA patches. If the sum over $\bk$ in Eq.\ref{eq:conduct} is carried out only in a single patch,  we can 
% obtain the antinodal/nodal component of the dc resistivity (to the leading order in the general Sommerfeld expansion),

% \begin{equation}
% \rho_T = -\frac{\im \Sigma(k_F, 0)}{\sum_{\bk}\big( \frac{\partial \epsilon_k}{\partial k_x} \big)^2 \delta \big(\mu-Re\Sigma(\bk,0) -\epsilon_{\bk} \big)}
% \end{equation}
% in above equation the real part of the self-energy $\re \Sigma(k,\omega)$ can in principle modify $\rho_{T}$.
% However, as shown in Supplementary Fig.~\ref{fig:ek}, $\re \Sigma(k_F,0)$ is essentially $T-$ independent and the effective Fermi level does not changes with $T$ at low temperatures in the NFL.  Therefore $\rho_{T}$ is proportional to $\im \Sigma(k_F, 0)$ or namely $\gamma_k(T)$, which leads to a $T-$ linear antinodal/nodal $\rho_{T}$ in the range $0.17 \lesssim p \lesssim 0.20$ for  $t^{\prime}/t = -0.2$.

\section{Temperature dependence of the effective dispersion $\epsilon^{*}(k)$ }\label{sup:ek}

We note that at small temperatures, the effective band dispersion
$\epsilon^{*}(k)=\epsilon(k)-\mu-\re \Sigma(\bk,\omega=0)$ is essentially $T-$ independent in the NFL, as shown in~\fref{fig:ek}. Therefore,
here the emergence of non-Fermi liquid properties,  such as the  non-saturating $\chi(T)$,  has nothing to do with a change of chemical potential or of quasiparticle number as $T$ changes~\cite{xu2013}.  Indeed, we have shown in the main text that  electrons in the NFL phase break Landau Fermi liquid theory in an intrinsic way, \textit{i.e.},  the electronic scattering rate  $\gamma_k \equiv -\mathrm{Im}\Sigma(k, \omega=0)$ disobeys the $T^{2}$ law of Fermi liquids.
 Another consequence of $\epsilon^{*}(k)$ being $T-$ independent is that the dc resistivity neglecting vertex corrections  in a homogeneous system $\rho_{T} =  6 \im \Sigma(k_F, 0)/e^2 {v_F}^2 N(0)$ ~\cite{Varma2020} would be proportional to the  scattering rate $ -\im \Sigma(k_F, 0)$. This is because the bare Fermi velocity $v_F$ and the bare density of states at the Fermi level $N(0)$  become
 constants when $\epsilon^{*}(k)$ is $T-$ independent. 

\begin{figure}
  \begin{center}
    \includegraphics[width=\columnwidth]{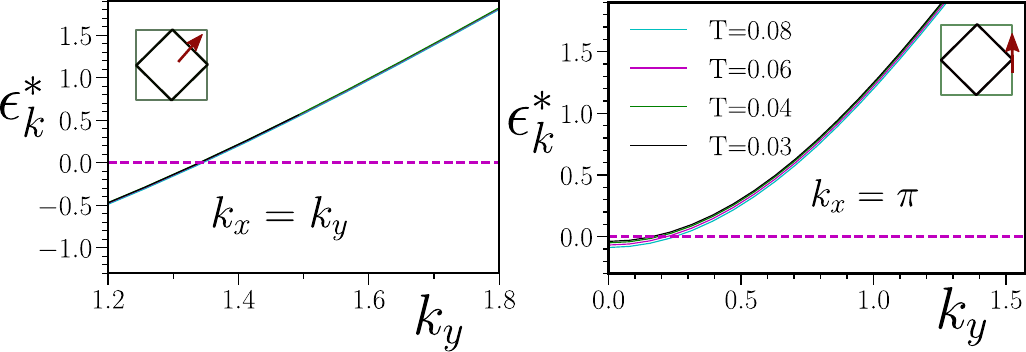}
  \end{center}
  \caption{
    \textbf{Effective dispersion $\epsilon^{*}(k)$ as a function of $\bk=(k_x, k_y)$ in two different directions in momentum $k$ space.  }
    \textbf{(a):} $\epsilon^{*}(k)$  in the nodal direction.
    \textbf{(b):}  $\epsilon^{*}(k)$ in the antinodal direction. Inserts indicate the cut taken in the Brillouin zone. 
    Here $p=0.18, U =7t, t^{\prime} = -0.2t$.
    }
\label{fig:ek}    
\end{figure}

\section{ Fluctuation analysis on selfenergy  in charge and particle-particle channel} \label{sup:fluc}

It has been shown that one can use the Dyson-Schwinger equation of motion (DSEOM) to decompose selfenergy at the two-particle level~\cite{gunnarsson2015, wu2017}, 
\begin{equation}
\Sigma(k)  =\frac{Un}{2}- \frac{U}{\beta^2N}\sum_{k^{\prime},Q}F_{\uparrow\downarrow}(k,k^{\prime},Q)g(k^{'})g(k^{\prime}+Q)g(k+Q)\\
\end{equation}
(with $Un/{2}$ the Hartree shift) in terms of the full two-particle scattering amplitude $F_{\uparrow\downarrow}(k,k^{\prime},Q)$.  The  full two-particle scattering amplitude $F_{\uparrow\downarrow}(k,k^{\prime},Q)$ can be rewritten in different sectors: spin (sp), charge (ch), or particle-particle (pp). For the Hubbard model,
\begin{eqnarray}
 F_{sp}(k,k^{\prime},Q)=F_{\uparrow\uparrow}(k,k^{\prime},Q)-F_{\uparrow\downarrow}(k,k^{\prime},Q)\\
 F_{ch}(k,k^{\prime},Q)=F_{\uparrow\uparrow}(k,k^{\prime},Q)+F_{\uparrow\downarrow}(k,k^{\prime},Q)\\
 F_{pp}(k,k^{\prime},Q)=F_{\uparrow\downarrow}(k,k^{\prime},Q-k-k^{\prime})
\end{eqnarray}
where $F_{\uparrow\uparrow}(k,k^{\prime},Q)=F_{\uparrow\downarrow}(k,k^{\prime},Q)-F_{\uparrow\downarrow}(k,k+Q,k^{\prime}-k)$.~\cite{gunnarsson2015}.
Hence for the DSEOM decompositions in different sectors,  $\Sigma_{sp/ch/pp}^{Q}(k)$, we have
\begin{eqnarray}
 \Sigma_{sp}^{Q}(k)= \frac{U}{\beta^2N}\sum_{k^{\prime}} F_{sp}(k,k^{\prime},Q)g(k^{'})g(k^{\prime}+Q)g(k+Q) \nonumber \\ 
 =\frac{U}{\beta^2N}\sum_{k^{\prime}}-F_{\uparrow\downarrow}(k,k+Q,k^{\prime}-k)g(k^{\prime})g(k^{\prime}+Q)g(k+Q)\\\label{eq:sp}
  \Sigma_{ch}^{Q}(k)= \frac{U}{\beta^2N}F_{ch}(k,k^{\prime},Q)g(k^{'})g(k^{\prime}+Q)g(k+Q) \nonumber \\
 =\frac{U}{\beta^2N}\sum_{k^{\prime}}[F_{\uparrow\downarrow}(k,k+Q,k^{\prime}-k)-2F_{\uparrow\downarrow}(k,k^{\prime},Q)]g(k^{\prime})g(k^{\prime}+Q)g(k+Q)\\
    \Sigma_{pp}^{Q}(k)= \frac{U}{\beta^2N}F_{pp}(k,k^{\prime},Q)g(k^{'})g(k^{\prime}+Q)g(k+Q) \nonumber \\
 =\frac{U}{\beta^2N}\sum_{k^{\prime}}F_{\uparrow\downarrow}(k,k^{\prime},Q-k^{\prime}-k)g(k^{\prime})g(k^{\prime}+Q)g(k+Q)
 \end{eqnarray}
 Note that the following sum-rule always hold for all the decompositions in different channels  $\Sigma_{sp/ch/pp}^{Q}(k)$,
 
\begin{equation}
\sum_{Q}(\Sigma_{sp,ch,pp}^{Q}(k))=\Sigma(k)-\frac{Un}{2}
\end{equation}
 
In practice, one does not need to explicitly compute the two-particle quantity $F_{\uparrow\downarrow}(k,k^{\prime},Q)$ and $ F_{\uparrow\uparrow}(k,k^{\prime},Q)$, then perform convolutions with Green's functions $g(k)$ to get self-energy decomposition $\Sigma_{sp/ch/pp}^{Q}(k)$ according to above equations.
For example, to obtain the self-energy decomposition in the \textit{spin}
channel $\Sigma_{sp}^{Q}(k)$, we can use Eq.~\ref{eq:sp}, the insert of Fig.~\ref{fig:sigma} and the notation  $k=(\bk, i\omega_n)$, $Q=(\bq, i\Omega_n)$, $R=(\br, \tau)$, to find,
\begin{eqnarray}
\Sigma^{Q}(k)+\frac{Un}{2}  =\frac{-U}{g(k)N^2\beta^2}\sum_{k^{\prime}} \langle S^{+}_{k}(-Q) S^{-}_{k^{\prime}}(Q) \rangle  \nonumber \\
=\frac{-U}{g(k)N^2\beta^2}\sum_{k^{\prime}}\sum_{(R_1, R_2,R_3,R_4)}  \langle C^{\dagger}_{\uparrow R_1}   C_{\uparrow R_2} C^{\dagger}_{\downarrow R_3} C_{\downarrow R_4} \rangle  \nonumber  
 \times e^{ikR_1} e^{-ik^{\prime}R_2} e^{i(k^{\prime}+Q)R_3} e^{-i(k+Q) R_4}  \\
 = \frac{-U}{g(k)N^2\beta^2}\sum_{(R_1, R_2,R_3,R_4)} \langle C^{\dagger}_{\uparrow R_1}   C_{\uparrow R_2} C^{\dagger}_{\downarrow R_3} C_{\downarrow R_4} \rangle 
 \times e^{ikR_1} e^{iQ(R_3-R_4)} e^{-ikR_4} \delta_{R_2, R_3} \nonumber \\ 
 =  \frac{-U}{g(k)N\beta}\sum_{(R_1, R_4)} \langle C^{\dagger}_{\uparrow R_1}   C_{\uparrow R_O} C^{\dagger}_{\downarrow R_O} C_{\downarrow R_4} \rangle 
 \times e^{ikR_1}  e^{-i(k+Q)R_4}
  \end{eqnarray}
where translational sysmetry  was used.  Here  $R_O$ is the original point in time and real-space, $R_O\equiv[\br_0=(0,0),\tau_0=0]$. Hence for the transfer momentrum $\bq$ decomposition of the self-energy at the first Matsubara 
frequency,  $\Sigma_{sp}^{\bq}(\bk, i\omega_0)$ reads,

%\begin{eqnarray}
\begin{align}
\Sigma_{sp}^{\bq}(\bk, i\omega_0) =  \sum_{\Omega_n} \Sigma^{\bq, \Omega_n}_{sp}(\bk, i\omega_0) \nonumber & \\
 =  \frac{-U}{g(k)N\beta}\sum_{\Omega_n}\sum_{(r_1, r_4)} \int_{\tau_1, \tau_4} \langle C^{\dagger}_{\uparrow \br_1}(\tau_1)   C_{\uparrow R_O} C^{\dagger}_{\downarrow R_O} C_{\downarrow \br_4}(\tau_4) \rangle  
 \times e^{i\bk \cdot \br_1}  e^{i\omega_0 \tau_1} e^{-i(\bk+\bq)\cdot \br_4} e^{-i(\omega_0+\Omega_n)\tau_4} \nonumber  \\
=  \frac{-U}{g(k)N}\sum_{(\br_1, \br_4)} \int_{\tau_1} \langle C^{\dagger}_{\uparrow \br_1}(\tau_1)   C_{\uparrow \br_0} (0)C^{\dagger}_{\downarrow \br_0}(0) C_{\downarrow \br_4}(0) \rangle   \times e^{i\bk \cdot \br_1} e^{-i(\bk+\bq)\cdot \br_4} e^{i\omega_0 \tau_1} d\tau_1  &
\end{align}
 %\end{eqnarray}
 while for the frequency decomposition in the \textit{spin} channel, we have, 
 \begin{align}
\Sigma_{sp}^{\Omega_n}(\bk, i\omega_0) =  \sum_{\bq} \Sigma^{\bq, \Omega_n}_{sp}(\bk, i\omega_0)  \nonumber \\
   =  \frac{-U}{g(k)N\beta}\sum_{\bq}\sum_{(\br_1, \br_4)} \int_{\tau_1, \tau_4} \langle C^{\dagger}_{\uparrow \br_1}(\tau_1)   C_{\uparrow R_O} C^{\dagger}_{\downarrow R_O} C_{\downarrow \br_4}(\tau_4) \rangle  
 \times e^{i\bk \cdot r_1}  e^{i\omega_0 \tau_1} e^{-i(\bk+\bq)\br_4} e^{-i(\omega_0+\Omega_n)\tau_4} \nonumber  \\
 =  \frac{-U}{g(k)\beta}\sum_{\br_1} \int_{\tau_1, \tau_4} \langle C^{\dagger}_{\uparrow \br_1}(\tau_1)   C_{\uparrow \br_0} (0)C^{\dagger}_{\downarrow \br_0}(0) C_{\downarrow \br_0}(\tau_4) \rangle   \nonumber  \\ 
 \times e^{i\bk \cdot \br_1}  e^{i\omega_0 \tau_1} e^{-i(\omega_0+\Omega_n)\tau_4}d\tau_1 d\tau_4
 \end{align}
 
Therefore for the DSEOM decomposition in the \textit{spin} channel, we only need to measure four-fermion correlators like  $\langle C^{\dagger}_{\uparrow \br_1}(\tau_1)   C_{\uparrow \br_0} (0)C^{\dagger}_{\downarrow \br_0}(0) C_{\downarrow \br_4}(0) \rangle $, which is similar to measuring the double occupancy $D_{occ}=\langle C^{\dagger}_{\uparrow}(0)C_{\uparrow}(0)C_{\downarrow}^{\dagger}(0)C_{\downarrow}(0)\rangle$.

For the decompositions in the \textit{charge} and \textit{particle-particle} channels, one can do similar derivations. For example, for the decomposition in the transfer momentum $\bq$ space,

\begin{eqnarray}
 \Sigma_{\uparrow \downarrow}^{\bq}(\bk, i\omega_0) 
 =  \frac{-U}{g(k)N}\sum_{(r_1, r_2)} \int_{\tau_1} \langle C^{\dagger}_{\uparrow \br_1}(\tau_1)   C_{\uparrow \br_2} (0)C^{\dagger}_{\downarrow \br_0}(0) C_{\downarrow \br_0}(0) \rangle   \nonumber  \\ 
 \times e^{i\bk \cdot r_1} e^{-i(\bk+\bq)\cdot \br_2} e^{i\omega_0 \tau_1} d\tau_1 \nonumber \\
 \Sigma_{ch}^{\bq}(\bk, i\omega_0) = 2 \Sigma_{\uparrow \downarrow}^{\bq}(\bk, i\omega_0) - \Sigma_{sp}^{\bq}(\bk, i\omega_0) 
 \end{eqnarray}

\begin{eqnarray}
 \Sigma_{pp}^{\bq}(\bk, i\omega_0) =
   \frac{-U}{g(k)N}\sum_{(\br_1, \br_3)} \int_{\tau_1} \langle C^{\dagger}_{\uparrow \br_1}(\tau_1)   C_{\uparrow \br_0} (0)C^{\dagger}_{\downarrow \br_3}(0) C_{\downarrow \br_0}(0) \rangle   \nonumber  \\ 
 \times e^{i\bk \cdot (\br_1-\br_3)} e^{-i\bq \cdot \br_3} e^{i\omega_0 \tau_1} d\tau_1
 \end{eqnarray}

In Fig.~\ref{fig:fluc2} we show that, for a typical doping in the NFL, and our usual parameters $U=7t, t^{\prime} = -0.2, p=0.18$, 
there are no prominent modes in the \textit{charge} and \textit{ particle-particle} channels that can dominate electron scattering.
Therefore we conclude that only spin collective modes  can contribute significantly to electronic scattering in the NFL. 
%{\color{blue} The frequency decomposition  $\im\Sigma^{\Omega_n}_{sp/ch/pp}(\bk, i\omega_0)$ will be included soon...
%}

\begin{figure}
  \begin{center}
    \includegraphics[width=\columnwidth]{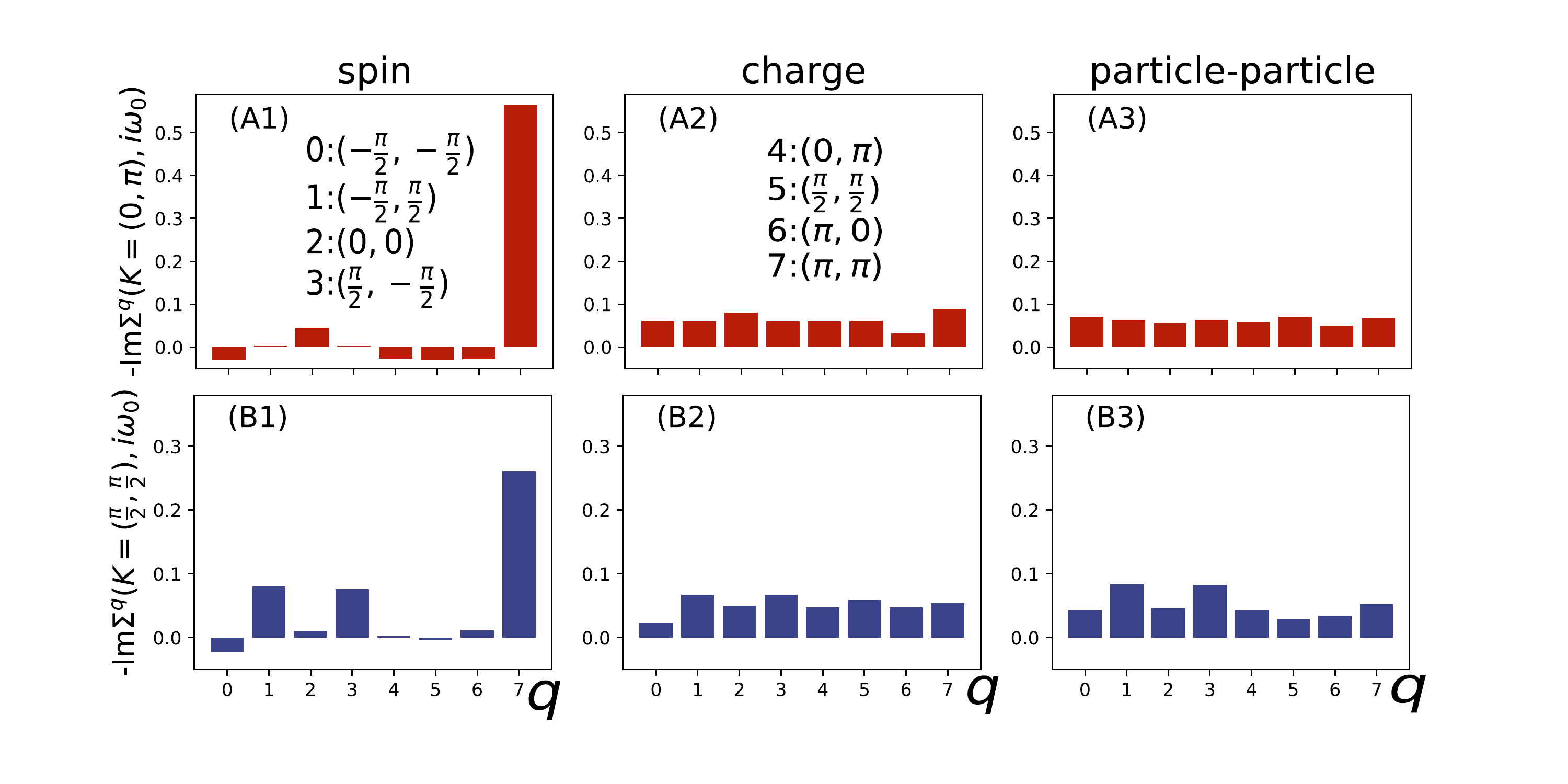}
  \end{center}
  \caption{
    \textbf{ Comparison of $\im\Sigma^{\bq}_{sp/ch/pp}(\bk, i\omega_0)$ at $U=7t, t^{\prime} = -0.2, p=0.18$ in three different channels: \textit{spin}, \textit{charge}, \textit{particle-particle}.   }
    \textbf{A1-A3:} For the antinode, $\bk = (0,\pi)$. \textbf{B1-B3: } For the node, $\bk = (\pi/2, \pi/2)$.
    }
\label{fig:fluc2}    
\end{figure}

\section{ Fluctuation analysis of the self-energy  at large dopings in the NFL} \label{sup:fluc2}

In the main text, we have shown $-\im\Sigma^{\bq/\Omega_n}_{sp}(\bk, i\omega_0)$ for $p=0.18$ and $p=0.20$ in the NFL.
In the following figure, we present the DSEOM decomposition in the \textit{spin} channel for more NFL doping levels. 

As we can see in Fig.~\ref{fig:fluc3}, for $p=0.24$ and $p=0.26$
at the antinode, $\bk=(0,\pi)$, the AFM $\bq=(\pi,\pi)$ fluctuations always have by far the  largest contribution to $-\im\Sigma(\bk, i\omega_0)$. 
For the node, $\bk=(\pi/2,\pi/2)$,
the $\bq=(\pi, \pi)$ mode contribution is still the largest, but there are also $\bq=(\frac{-\pi}{2},\frac{\pi}{2})$,
$(\frac{\pi}{2},\frac{-\pi}{2})$ modes that lead to significant sources of
 scatterings. Looking carefully, for the node $\bk=(\frac{\pi}{2},\frac{\pi}{2})$,
scatterings from these two magnetic modes actually involve $(0,\pi)$/$(\pi,0)$ momenta in the Dyson Schwinger equation [since $\bk+\bq$=$(0,\pi)$/$(\pi,0)$ for $\bq=(\frac{-\pi}{2},\frac{\pi}{2})$,
$(\frac{\pi}{2},\frac{-\pi}{2})$ respectively]. Since $(0,\pi)$,$(\pi,0)$ are van Hove singularities (VHS), flatband effects can increase
scattering phase space. So
we argue that nodal electrons in the NFL can be scattered relatively more frequently by non-($\pi,\pi$) modes, given also that $(\pi,\pi)$ antiferromagntic correlations are suppressed by doping.
 (Note that for the antinode $\bk = (0,\pi)$ or  $\bk = (\pi,0)$, the  $\bq=(\pi,\pi)$ mode always scatters electrons between VHS, since $\bk, \bk+\bq$ are both VHS).

\begin{figure}
  \begin{center}
    \includegraphics[width=\columnwidth]{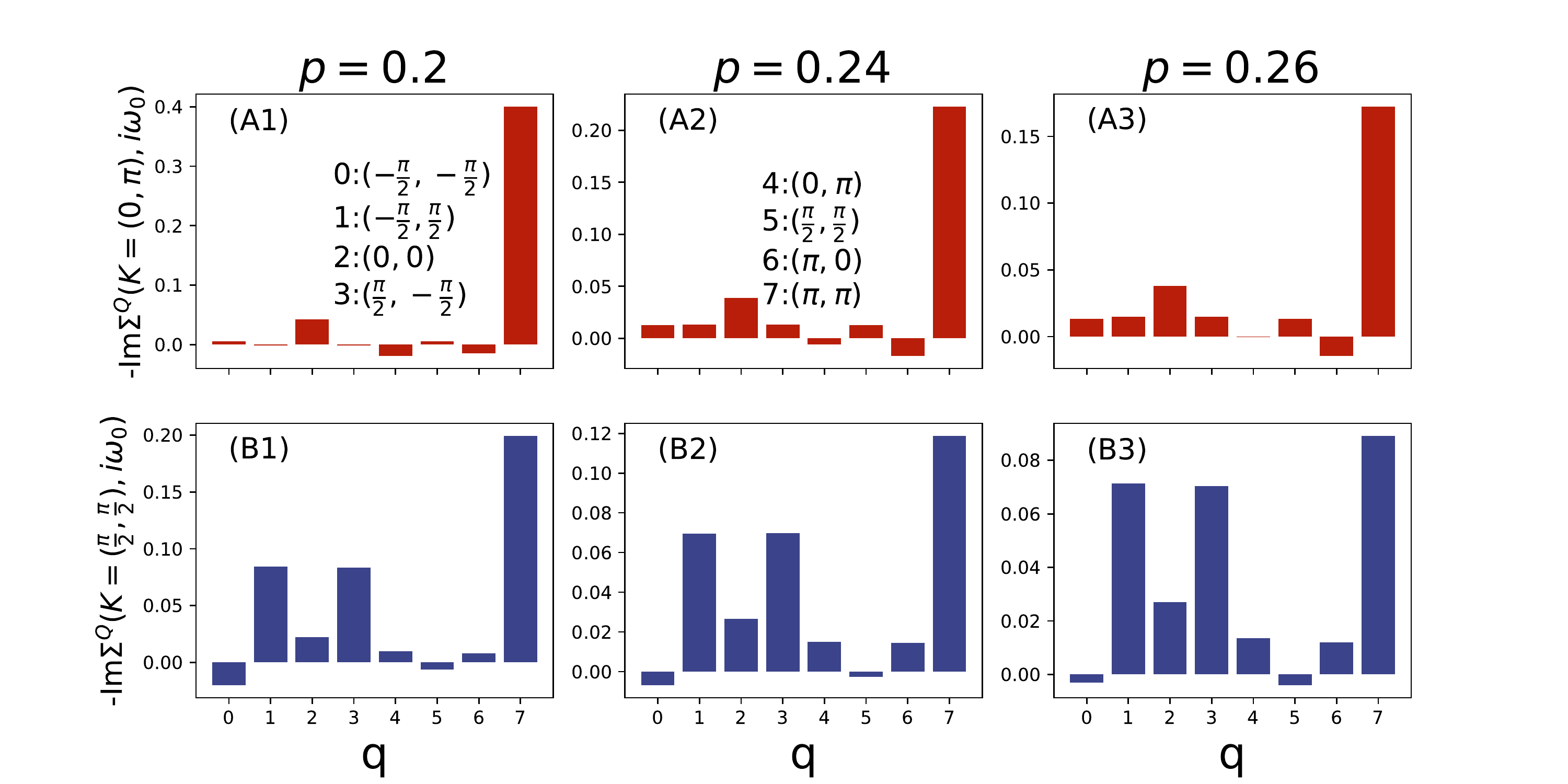}
  \end{center}
  \caption{
    \textbf{ Comparison of $\im\Sigma^{\bq}_{sp/ch/pp}(\bk, i\omega_0)$ at $U=7t, t^{\prime} = -0.2, p=0.18$ in the  \textit{spin} channel, for three different dopings }
    \textbf{A1-A3:} For the antinode, $\bk = (0,\pi)$. \textbf{B1-B3: } For the node, $\bk = (\pi/2, \pi/2)$.
    }
\label{fig:fluc3}    
\end{figure}

\section{ DCA cluster size effect:  Four-site and sixteen-site results} \label{sup:size}

Here we show results from larger $4\times 4$ DCA cluster computations. Owing to the minus sign problem of the impurity solver, we are not able to do calculations at temperatures $T$ as low as those for the 8-site cluster. As shown in Fig.~\ref{fig:gamma16} we are still able to obtain a $T-$ linear scattering rate up to relatively low-temperatures, namely $T/t \sim 0.1$ (at $p=0.05, 0.08$ in the underdoped regime,  when $T^{*}$ is not yet reached). The fluctuation analysis in the $T-$ linear regime also suggests that the AFM fluctuations  $\bq=(\pi, \pi)$ are the main source of $T-$ linear electronic scattering rate, as shown in Fig.~\ref{fig:gamma16}. For a smaller $2\times 2$ cluster we obtained the same results (not shown here). 

\begin{figure}
  \begin{center}
    \includegraphics[scale=0.6]{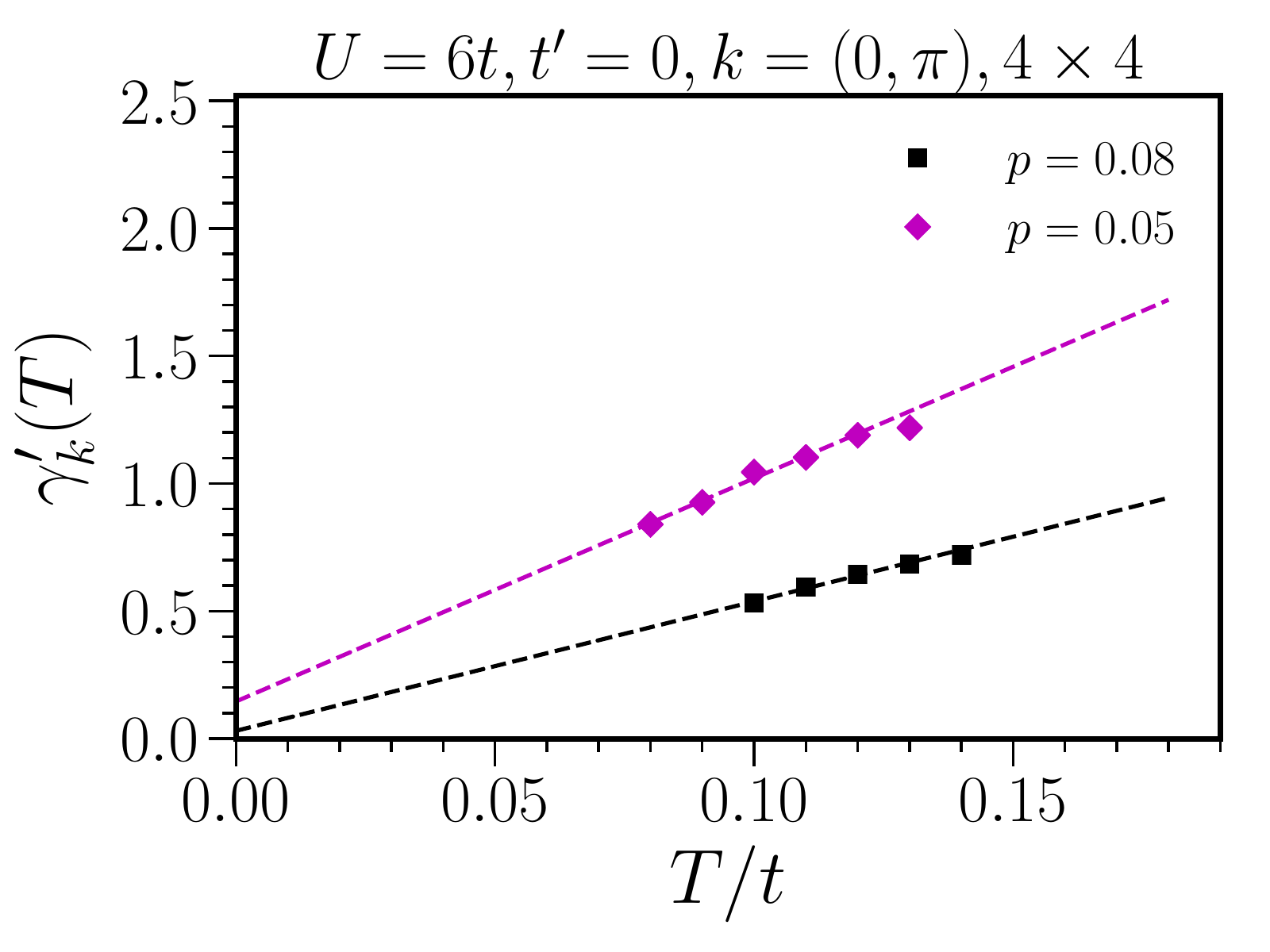}
  \end{center}
  \caption{
   Antinodal scattering rate $\gamma^{\prime}_k(T)$, $\bk=(0,\pi )$ as a function of temperature $T$ at two dopings for $4\times 4$ DCA cluster. 
    }
\label{fig:gamma16}    
\end{figure}

\begin{figure}
  \begin{center}
    \includegraphics[width=\columnwidth]{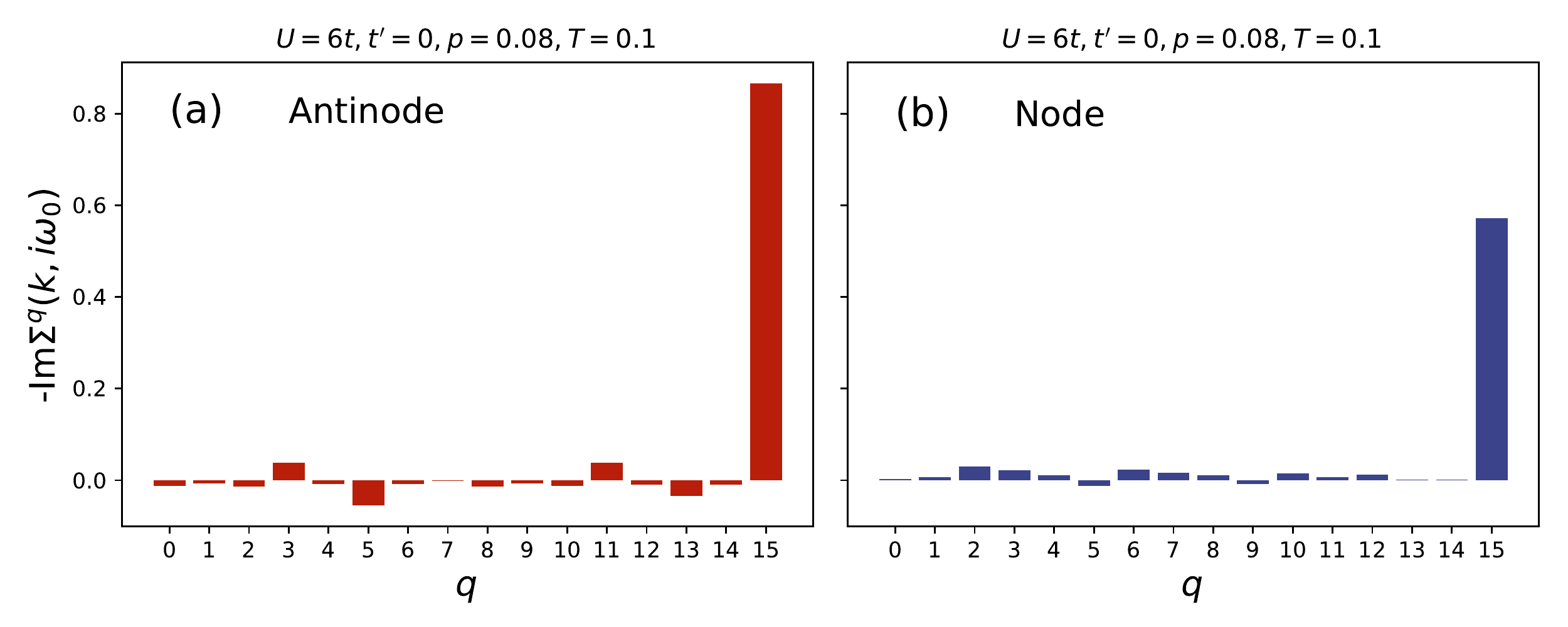}
  \end{center}
  \caption{
   Dyson-Schwinger equation of motion decomposition of the self-energy in the \textit{spin} channel, $\im\Sigma^{\bq}_{sp}(\bk, i\omega_0)$ for the $4\times4$ DCA cluster. 
The $\bq$ indices are,
0:$(-\frac{\pi}{2},-\frac{\pi}{2})$
1:$(-\frac{\pi}{2},0)$ 
2:$(-\frac{\pi}{2}, \frac{\pi}{2})$
3:$(-\frac{\pi}{2},\pi)$ 
4:$(0,-\frac{\pi}{2})$ 
5:$(0,0)$
6:$(0,\frac{\pi}{2})$
7:$(0,\pi)$
8:$(\frac{\pi}{2},-\frac{\pi}{2})$
9:$(\frac{\pi}{2},0)$
10:$(\frac{\pi}{2}, \frac{\pi}{2})$
11:$(\frac{\pi}{2}, \pi)$
12:$(\pi, -\frac{\pi}{2})$
13:$(\pi, 0)$
14:$(\pi, \frac{\pi}{2})$
15:$(\pi,\pi)$.
    }
\label{fig:fluc16}    
\end{figure}

\section{On the nearly Planckian liquid} \label{sup:dimensional}

We focused on the so-called strange metal, that refers to the regime where a linear temperature dependence of the scattering rate extends all the way to zero temperature. The case where the coefficient $C$ of the scattering rate $CT$ is equal to unity (in units $k_B=1$, $\hbar=1$) is conjectured in the literature to be a fundamental limit, the ``Planckian limit''\cite{Bruin_Sakai_Perry_Mackenzie_2013,zaanen2019,Legros2019,grissonnanche2020,hartnoll2021planckian}.   

A linear in $T$ scattering rate follows, for example, in the case of phonon scattering when $T$ is larger than the Debye frequency, because then the number of bosonic scatterers is proportional to $T$~\cite{Sadovskii_2021}. A similar idea has been proposed in the case of an antiferromagnetic QCP\cite{ millis1993, gegenwart2008,  lohneysen2007, Xu2020,dumitrescu2021} because at the QCP the characteristic spin fluctuation frequency, that plays a role analog to the Debye frequency, vanishes. The latter explanation does not hold in the weak correlation limit for two reasons. First, the QCP occurs at an isolated doping and, second, one expects that scattering will be strong only at hot spots on the Fermi surface so that the scattering rate will not be isotropic and, barring disorder effects~\cite{Rosch:1999}, the resulting resistivity will be short-circuited by Fermi-liquid-like portions of the Fermi surface~\cite{Hlubina:1995}.

The strong to intermediate correlation limit that we have considered here seems to solve the above two problems.
First, the linear $T$ NFL regime holds in a finite range of overdoping, as observed experimentally~\cite{cooper2009}.
Of course,  here we should leave open the possibility that the finite intercept found for the antinodal scattering rate could be a signature of finite crossover temperatures ($T^*$ or $T_{FL}$) that are too small to be accessible numerically ($ < 0.025t$) .
Clearly, however, our calculations strongly suggest that the extrapolated crossover temperatures are very small if not vanishing (see  Fig.~\ref{fig:allanti}).

Second, for the strong interaction, $U=7t$, that we considered, it is quite possible that the lack of well-defined quasiparticles leads to spin fluctuations with vanishing characteristic frequency. Then, the argument that the number of scatterers scales like $T$ should hold. Moreover, in over-doped regime  the correlation length is small~\cite{kastner1998} so that the spin fluctuations can  scatter effectively electrons at all the remains of the Fermi surface. Hence, the argument that the  T-linear scattering rate is isotropic will also hold.
%(Note that for the antinodal scattering in the pseudogap phase,  the proximity to van Hove singularities~\cite{wu2020} plays a vital role, thus the above analysis does not apply) .
The only question left then, is why is the coefficient close to unity for many materials. We offer the following explanation. On dimensional grounds, we can write, (restoring physical units)
\begin{equation}
  -\im\Sigma(T)=k_BT \times f(\frac{k_BT}{\hbar\omega^*},\frac{U_e}{W})
\end{equation}
where $f$ is a dimensionless scaling function while $\hbar\omega^*$ is the characteristic spin-fluctuation energy, $W$ the bandwidth and $U_e$ the screened interaction. When $\hbar\omega^*$ is large, a Taylor expansion of the scaling function in terms of its first argument gives the Fermi liquid result that the scattering rate is proportional to $(k_BT)^2$. Following the argument of Kanamori~\cite{kanamori_electron_1963} and Brückner~\cite{Brueckner:1960}, the bare interaction $U$ is screened by quantum fluctuations and the resulting screened interaction $U_e$ becomes nearly equal to the bandwidth $W$ in the dilute limit. Physically, when $U$ is large, the two-particle wave function tends to vanish when two electrons are on the same site. The maximum energy that this can cost is the bandwidth $W$, that becomes the effective interaction energy. While this result can be proven when $U$ is not too large, it is natural to assume that it holds here. In the limit where $\hbar\omega^*$ vanishes then, we have   
\begin{equation}
  -\im\Sigma(T)=k_BT \times f(\infty,a)
\end{equation}
where $a$ is a number of order unity for a wide range of bare $U$, following the Kanamori-Brückner argument. So $-\im\Sigma(T)$ can take similar values for a wide class of materials whose low-energy behavior is described by a Hubbard model. Since dimensionless functions are usually of order unity, this suggests why the prefactor of $k_BT$ is of order unity. But it clearly does not need to be unity. In addition, other dimensionless quantities can appear as additional arguments of this function, such as the ratio $t'/t$. In fact, we find a number about equal to three for this function. So we call the strong-interaction case that we studied, a ``nearly Planckian liquid'' and we argue that Planckian dissipation is not a fundamental limit to the electron scattering rate~\cite{Sadovskii_2021}.  

\end{widetext}
\end{document}